\documentclass[a4paper,twocolumn,11pt,accepted=2023-09-07]{quantumarticle}
\pdfoutput=1

\usepackage{braket}
\usepackage{graphicx}
\usepackage{dcolumn}
\usepackage{bm}
\usepackage{hyperref}
\usepackage[mathlines]{lineno}
\usepackage{subcaption}
\usepackage{amsmath,amssymb}
\usepackage[ruled]{algorithm2e}
\SetKwComment{Comment}{/* }{ */}
\usepackage{soul}

\begin{document}

\newcommand{\roundd}[2]{{#2}}


\title{Coherent errors and readout errors in \roundd{}{the} surface code}

\author{\'A. Márton}
\affiliation{Department of Theoretical Physics, Institute of Physics,
Budapest University of Technology and Economics, Műegyetem rkp. 3., H-1111 Budapest, Hungary}%
\author{J. K. Asb\'oth}
\affiliation{Department of Theoretical Physics, Institute of Physics,
Budapest University of Technology and Economics, Műegyetem rkp. 3., H-1111 Budapest, Hungary}%
\affiliation{Wigner Research Centre for Physics, H-1525 Budapest, P.O. Box 49., Hungary}


\begin{abstract}
We consider the combined effect of readout errors and coherent errors, i.e., deterministic phase rotations, on the surface code.
We use a recently developed numerical approach, via a mapping of the physical qubits to Majorana fermions. 
We show how to use this approach in the presence of readout errors, treated on the phenomenological level: perfect projective measurements with potentially incorrectly recorded outcomes, and multiple repeated measurement rounds. 
We find a threshold for this combination of errors, with an error rate close to the threshold of the corresponding incoherent error channel (random Pauli-Z and readout errors). The value of the threshold error rate\roundd{ depends on how the logical-level errors are quantified: using the diamond norm, it is 3.1\%, using for logical-level fidelity, it is 2.6\%.}
{, using the worst case fidelity as the measure of logical errors, is 2.6\%.} 
Below the threshold, scaling up the code leads to the rapid loss of coherence  in the logical-level errors, but error rates that are greater than those of the corresponding incoherent error channel. We also vary the coherent and readout error rates independently, and find that the surface code is more sensitive to coherent errors than to readout errors. Our work extends the recent results on coherent errors with perfect readout to the experimentally more realistic situation where readout errors also occur.     
\end{abstract}

\maketitle

\section{Introduction}

The surface code\cite{kitaev_preskill2002, fowler2012surface} is one of the most promising candidates for quantum error correction. 
For a code patch of distance $d$, the collective quantum state of $d^2$ physical qubits is used to store a single logical qubit.
Incoherent errors (from entanglement of the physical qubits with a memoryless environment) can be modeled as random Pauli operators on the physical qubits. 
Repeated measurements of the parity check operators (a.k.a. stabilizer generators, to obtain the so-called syndrome) 
can be used to localize and correct such errors, with a success probability that increases as the code distance increases, $d\to \infty$, as long as the error rates are below the so-called threshold. 

For incoherent errors, the value of the threshold depends on the details of the error model, of the error correction (decoding) procedure, and on the level of detail in which the measurements are modeled (circuit-level or not). 
For simple cases, the threshold is known from mappings between the correction of incoherent errors on the surface code and phase transitions in classical Ising models\cite{kitaev_preskill2002, wang2003confinement}: it is around 10\% with perfect and around 3\% with imperfect measurements. This mapping can be extended to some other error models and codes as well\cite{bombin2012strong,chubb2021statistical}.
For more complicated error models and circuit-level modeling the threshold can be obtained numerically, using efficient simulation in the Heisenberg picture\cite{aaronson2004improved, gidney2021stim}, and is around 0.75\%. These threshold values not very far from current experimental reality: below 1\% for quantum gates, and a few \% for readout\cite{krinner2022realizing, google2023suppressing}. 



The effect of coherent errors on the surface code is less well understood. These are errors modeled by nonrandom unitary operators acting on each physical qubit separately at each timestep. In the simplest case -- the one we will also consider -- these are phase rotations of the qubits with a fixed angle $\theta$, i.e., $e^{i\theta\hat{Z}}$. 
Such errors model the effect of components becoming miscalibrated, inevitable in long calculations.
Since coherent errors are not Clifford operations, their effects cannot be simulated efficiently in the Heisenberg picture\cite{aaronson2004improved}. 
Brute-force simulations\cite{tomita_svore, greenbaum_dutton}, tensor network methods\cite{tensor_network} or other efficient approximations\cite{hakkaku2021sampling}, 
and the standard mappings to statistical physics models also break down (but note a  recent extension of the mappings to a Majorana scattering network\cite{venn2022coherent}).

One of the ways in which coherent errors are more complicated than incoherent errors is that they lead to coherent errors on the logical level as well. With coherent errors, the parity check measurements project the surface code into a random state, which even after error correction differs from the original state. For codes with an odd distance, this difference corresponds to a coherent rotation on the logical level. This logical-level coherence also makes the quantitative investigation of a quantum memory more complicated. Note, however, that scaling up the surface code leads to a ”washing out” of coherence from the errors on the logical level. As
shown analytically\cite{beale2018quantum, iverson_preskill}, the random logical-level rotations correspond more and more to random Pauli noise as the code size is increasing. 
Note also, however, that the rate of this logical-level Pauli noise has not been computed analytically, and it can be considerably higher than what one would get by simply replacing coherent errors with their Pauli twirled counterparts\cite{gutierrez2016errors, greenbaum_dutton}. 

Coherent errors also seem to have a threshold, as shown numerically by \roundd{Bravy}{Bravyi} et al\cite{bravyi2018correcting}. They have simulated relatively large code sizes using a mapping to Majorana fermions. They have seen the "washing out \roundd{}{of} the coherence", and found that the rate of the logical-level Z error is \roundd{indeed}{}  significantly higher than \roundd{those}{that} obtained by Pauli twirling the physical-level errors. Nevertheless, their numerics revealed an error threshold for coherent errors too: for $\theta < \theta_{th}\approx 0.08\pi-0.1\pi$, the logical-level error rates decrease as the code is scaled up. This value of the threshold is very close to that of the  Pauli twirled physical error channel: $\sin(\theta_{th})^2\approx 0.09$. 

In this paper we bring the results on coherent errors closer to experimentally relevant setting by considering them together with measurement errors. We consider the simplest kind of measurement errors, the so-called phenomenological error model: perfect projective von Neumann measurements of the parity check operators, with possibly incorrectly recorded measurement results. We use the numerical simulation approach based on the mapping to Majorana fermions\cite{bravyi2018correcting, benjamin_florian}, but combine this with readout errors, and a corresponding 3D decoding.  


This paper is structured as follows. 
In Sec.~\ref{sec:Surface_code} we briefly introduce the most important concepts of the surface code, including error correction by minimum weight perfect matching on a 3D syndrome graph. In Sec.~\ref{sec:Simulation_method} we introduce the key points of the simulation method using fermionic linear optics, as pioneered by Bravyi et al\cite{bravyi2018correcting}. In Sec.~\ref{sec:Results} we present our results on the combined effects of coherent and readout errors on the surface code. In Sec.~\ref{sect:Discussion} we conclude the paper with a discussion of our results.

\section{Surface code} 
\label{sec:Surface_code}

We briefly introduce the surface code as a quantum memory, storing a single logical qubit, in the rotated basis\cite{tomita_svore}, with the patch encoding\cite{bombin2007optimal}.

\subsection{Definition of the code space}

\begin{figure}[!ht]
    \centering
    \includegraphics[width=.4\textwidth]{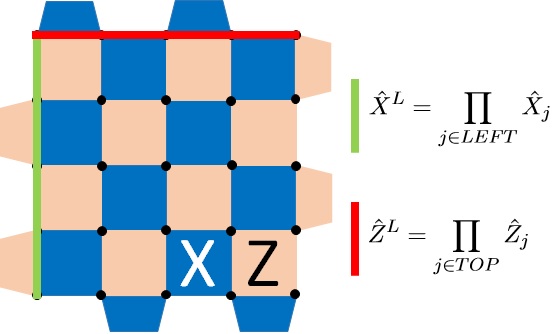}
    \caption{A patch of a surface code with code distance $d=5$. }
    \label{fig:surface_code}
\end{figure}

A distance-$d$ patch of the surface code (we always take $d$ odd) consists of $n=d^2$ physical (data) qubits arranged in a square grid. An example with $d=5$ is
shown in Fig.~\ref{fig:surface_code}.  The faces of the grid are
colored in a checkerboard pattern, light (brown) and dark (blue).
Extra boundary faces are also included, on all the edges, to ensure
that top and bottom edges consist of only blue faces, while left and
right edges of only light faces (smooth/rough
boundaries\cite{fowler2012surface}).

All faces of the grid correspond to parity check operators, which are the stabilizer generators of the code. 
These are products of Pauli operators on the qubits at the corners of the corresponding face.  
For each light (dark) face, $\hat{Z}$ ($\hat{X}$) operators are used, i.e.,
\begin{align}
    \hat{A}_f&=\prod_{j \in \partial f}\hat{X}_j;& 
    \hat{B}_f&=\prod_{j \in \partial f}\hat{Z}_j;\\
    \hat{S}_f&=\hat{A}_f \text{ if }  f\text{ 
    dark};& 
    \hat{S}_f&=\hat{B}_f \text{ if }  f\text{ light} .
\end{align}
They have eigenvalues $\pm 1$, which correspond to even/odd parity of
the corresponding group of qubits (in $X$ or $Z$ basis).  Since any \roundd{
two faces share either no corners or two corners}
{dark face shares either no corners or two corners with any light face}, all of these stabilizer operators commute.

The code space ("quiescent state"\cite{fowler2012surface})
of the surface code is defined as the +1 eigenspace of all the
stabilizers. This is a two-dimensional space spanned by the logical
basis states,
\begin{align}
    \ket{0_L} &= N_d \prod_{f\in\text{blue}} \dfrac{1}{2} (1+\hat{A}_f) \ket{0}^{\otimes n};\\
    \ket{1_L} &= N_d \prod_{f\in\text{blue}} \dfrac{1}{2}  (1+\hat{A}_f) \ket{1}^{\otimes n},
\end{align}
where $\ket{0}^{\otimes n}$ and $\ket{1}^{\otimes n}$ are the states
where all physical qubits are $\ket{0}$ and $\ket{1}$ respectively,
and 
$N_d= 2^{(d^2-1)/4}$
is a normalizing factor.  
Both logical basis states are highly entangled states of the physical qubits, and are locally indistinguishable from each other (all physical qubits have completely mixed \roundd{}{reduced}  density matrices). 
The encoding of a logical qubit is a mapping from a Hilbert space of dimension 2 to one of dimension $2^n$,
\begin{align}
    \ket{\psi}=\alpha \ket 0 + \beta \ket 1 \to 
    \ket{\psi_L}=\alpha\ket{0_L} + \beta\ket{1_L}.  
\end{align}

The logical (or "encoded") operators $\hat{X}^L$ and $\hat{Z}^L$ must
fulfil $\hat{X}^L \ket{0_L}=\ket{1_L}$, $\hat{X}^L
\ket{1_L}=\ket{0_L}$, $\hat{Z}^L
\ket{0_L}=\ket{0_L}$, $\hat{Z}^L
\ket{1_L}=-\ket{1_L}$.  One possible choice of such
logical operators is products of $\hat{X}$, ($\hat{Z}$)
on qubits along the left (top) edge,
\begin{align} \label{eq:logical operators}
    \hat{X}^L&=\prod_{j\in \text{LEFT}} \hat{X}_j;&  
    \hat{Z}^L&=\prod_{j\in \text{TOP}} \hat{Z}_j,
\end{align}
as shown in Fig.~\ref{fig:surface_code}.

    
\subsection{Coherent errors and their detection}

Of the various error processes by which the environment can affect the states of the physical qubits, we focus on coherent errors\cite{bravyi2018correcting}.  
Here each physical qubit undergoes a fixed SU(2) unitary operation in every timestep, resulting from e.g. calibration errors in a quantum computer. 
For simplicity, we take this unitary to be the same for every qubit, specifically, a rotation about the Z axis through an angle $\theta$ (noise parameter)\cite{bravyi2018correcting}. 
Thus the unitary operator representing the effect of noise reads,
\begin{align} \label{eq:coherent errors}
    \hat{U}=\prod_{j=1}^n e^{i \theta \hat{Z}_j}.
\end{align}
The noise parameter $\theta$ can be converted to a \emph{physical
error rate} $p$, as 
\begin{align}
    p=\sin^2(\theta).
\end{align}
This is the parameter of the dephasing channel obtained by Pauli twirling the coherent errors.

To detect and correct the coherent errors, we use the standard
procedure of repeated
measurements of the parity check operators.
Each measurement results in a measurement outcome and a
post-measurement quantum state. The string $s$ of measurement outcomes
is the \emph{syndrome},
\begin{align}
    \text{syndrome } s = (s_1, \ldots, s_{n-1}),
\end{align}
with all elements $s_f \in \{+1, -1\}$.
The (unnormalized) post-measurement state of the code can be obtained by a projection of the pre-measurement state, 
\begin{align}
    \hat{\Pi}_s &= \prod_{\forall f} \dfrac{1}{2} (1+s_f\hat{S}_f);& \ket{\Phi_s} &= \hat{\Pi}_s \ket{\Phi}.
\end{align}
Note that since we consider only coherent errors that are
$Z$-rotations, only the $X$-parity check measurements can return with
a $-1$ value.

\subsection{Error correction with perfect readout}

Before we discuss readout errors, we need to briefly summarize how the errors would be corrected if readout was perfect. 

If some of the parity check measurements have resulted in an outcome
of -1, a \emph{correction operation} $\hat{C}_s$ is needed to bring the state back into the code space. 
Since the coherent errors only contain $\hat{Z}$, this
correction involves flipping some qubits in the $X$ basis by
$\hat{Z}$, i.e., 
\begin{align}
    \hat{C}_s &= \prod_{j\in l}\hat{Z}_j.
\end{align}
The set $l$ of qubit indices is in a properly defined sense a 1-chain that connects the error locations with each other or with the left or right edge \cite{kitaev_preskill2002}, i.e., whose boundary (in a homological sense) is the set of faces where the measured parity check operators are $-1$.

Deciding on the correction operator for a given syndrome is the \roundd{solution of the }{}decoding problem. 
There are many possible correction operators for any given syndrome, which fall into two homological equivalence classes: when multiplying \roundd{}{any} two \roundd{}{correction} operators, if they are in the same class, we obtain a product of stabilizers, if they are in different classes, we obtain a logical $\hat{Z}^L$ times a product of stabilizers.  
In principle, the likelihoods of the two classes (based on the error model) should be compared and any correction operator from the more likely class should be chosen.  
However, given the computational cost of the likelihood calculation, various approximate approaches, so-called decoders, have been developed \cite{delfosse2021almost, bravyi2014efficient, MWPM}. 
We will discuss this in more detail after the introduction of readout errors. 

The collective quantum state of $n$ qubits after we measured syndrome $s$, and applied the corresponding correction operator, reads
\begin{align} \label{eq:final_coherent_start}
    \ket{\Phi_s}=\dfrac{1}{\sqrt{P(s)}}\hat{C}_s
    \hat{\Pi}_s\hat{U}\ket{\psi_L}.
\end{align}
\roundd{This is in the code subspace. Moreover, because the coherent errors are only Z-rotations (and the code distance is odd cite[bravyi2018correcting]), this differs from the original state only by a logical Z-rotation:}
{This state is in the logical subspace. Moreover it can be written as a rotation around the logical Z axis\cite{bravyi2018correcting} with the \emph{logical rotation angle} $\theta_L(s)$,}
\begin{align} 
\label{eq:final_state_coherent}
   \ket{\Phi_s}=e^{i\textrm{\roundd{$\theta_s$}{$\theta_L(s)$}} 
   \hat{Z}^L}\ket{\psi_L}.
\end{align}
\roundd{Interestingly the logical rotation angle $\theta_s$ depends on the syndrome, but not on the initial state $|\psi_L>$ cite[bravyi2018correcting]; this is not the
case for the surface code on some other types of
lattices cite[benjamin-florian]. Moreover, the probability $P(s)$ of
measuring the syndrome $s$ is also independent of the initial state $|\Psi_L>$ cite[bravyi2018correcting].}
{Here both the probability $P(s)$ and the angle $\theta_L(s)$ are independent of the initial state $\ket{\Psi_L}$. These are unique properties of the rotated surface code, and are not necessarily true for surface codes on general lattices\cite{benjamin_florian}. To obtain these properties all the X-stabilizers need to have even weight, while logical Z-operators should have odd weight. 
If this is not fulfilled, Eq.~\eqref{eq:final_coherent_start} can not be written as a logical rotation around the Z-axis, furthermore, the syndrome measurement is a weak measurement of the initial logical state $\ket{\Psi_L}$, resulting in an information "leak"\cite{huang2019performance}.}

\subsection{Quantifying logical errors}

To characterize the effectiveness of error correction, we use two quantitative measures, the \emph{diamond-norm distance} of a channel to the identity, and the \emph{maximum infidelity} of the resulting state with the initial state\cite{gilchrist2005distance}.
These show how different the state of the encoded qubit is after the error correction process from the original state.
In our case the average over syndromes of the diamond-norm distance can be expressed as \cite{bravyi2018correcting}:
\begin{align}
    p_L^d=2\sum_s P(s)|\sin(\theta_L(s))|,
\end{align}
and the average over the syndromes of the maximum infidelity as
\begin{align}
    p_L^i=\sum_s P(s)\sin^2(\theta_L(s)).
\end{align}
\roundd{}{Of these two measures, we prefer the maximum infidelity, since it is a more natural generalization of the logical error rate for coherent errors. 
However, we have also calculated the diamond-norm distance, and used it for the numerics - please find the corresponding analysis in   Appendix~\ref{apx:diamond_norm}.}

\subsection{Readout errors, 3D syndrome}

We take into account not only coherent errors on the physical qubits, but also readout errors distorting the result of the syndrome measurements.  
We consider the simplest, \emph{phenomenological} noise model for the readout \cite{delfosse}\roundd{. 
Thus, we assume}{:} perfect syndrome measurements, whose outcome is unreliably recorded, with a readout error probability $q$, i.e.,
\begin{align}
  P(1\rightarrow0) = P(0\rightarrow1)=q,
\end{align}
and correspondingly, $P(1\rightarrow1) = P(0\rightarrow0)=1-q$.
The obtained noisy syndrome is
\begin{align}
  \text{noisy syndrome:} \quad s \to s'.
\end{align}

To solve the decoding problem in the presence of readout errors, we need to consider $d$ consecutive rounds of syndrome measurements\cite{fowler2012surface, kitaev_preskill2002}.
Since errors occur between the rounds of syndrome measurements, the rounds of measurement outcomes differ from each other even if the measurements are perfect.
The $d$ rounds of syndromes constitute a \emph{3D syndrome}, which  without/with readout errors is 
\begin{align}
    \underline{s}=\{s_1,s_2,...,s_d\} \to 
    \underline{s}'=\{s_1',s_2',...,s_d'\}.
\end{align}

For error correction we have to solve the decoding of this 3D syndrome with some decoding technique, obtaining a correction operator $\hat{C}_{\underline{s}'}$. We will detail the decoding in the next Section. 

We can express the final state of the code for a measured 3D syndrome $\underline{s'}$, where the corresponding 3D syndrome without readout error was ${\underline{s}}$, as  
\begin{align} \label{eq:final_state_start}
    \ket{\Phi_{\underline{s},\underline{s}'}} = 
    \dfrac{1}{\sqrt{P(\underline{s})}} \hat{C}_{\underline{s}'} \hat{\Pi}_{s_d} \hat{U} \ldots \hat{\Pi}_{s_1} \hat{U} \ket{\psi_L}.
\end{align}
Using $\hat{U}=\prod_{j=1}^n\exp({i\theta \hat{Z}_j})$, this can be rewritten (more details in the next section) as
\begin{align} \label{eq:final_state_coherent_readout}
    \ket{\Phi_{\underline{s},\underline{s}'}}
    =
    \hat{C}_{\underline{s}'} \hat{C}_{s_d}
    e^{i\theta^*(\underline{s}) \hat{Z}^L}\ket{\psi_L}.
\end{align}
Here the rotation angle $\theta^*(\underline{s})$ only depends on the noiseless 3D syndrome $\underline{s}$, but not on the readout errors, nor on the initial state $\ket{\psi_L}$.

\subsection{Error correction with readout errors}

To correct errors based on the noisy 3D syndrome contaminated with readout errors we used the 3D version of the minimum weight perfect matching (MWPM) decoder\cite{MWPM}, as implemented in PyMatching\cite{pymatching}.
In this 3D case, like in the case with perfect measurements, errors are associated with marked vertices on a grid, we need to find the set of edges on the grid with the smallest weight that pair the vertices up or connect them to the right/left boundaries. 

The grid here is 3-dimensional, with "space" coordinates ($d\times d$ grid) giving the position of the measured stabilizer operator and "time" coordinates (running from $2$ to $d$) corresponding to the measurement round. Those vertices are marked where the measured stabilizer value differs from that measured in the previous round. "Spacelike" and "timelike" edges on the grid correspond to readout errors and physical errors. 
These carry different weights ($w_s$ and $w_t$), since the rate $p$ of coherent errors can differ from the rate $q$ of readout errors, 
\begin{align}
    w_s&=\log\left(\dfrac{1-p}{p}\right);& w_t&=\log\left(\dfrac{1-q}{q}\right).
\end{align}

The MWPM decoder finds the set of edges with the smallest overall weight, which perfectly connect the marked vertices (with each other, or with the left or right boundaries).
The set of edges with the minimum weight is used to define the correction operator, which consists of a string of $\hat{Z}$ operators. Spacelike edges correspond to $\hat{Z}$ operators, however timelike edges have no physical meaning, they are just virtual corrections of readout errors. 
A technical note: to ensure the correction operator brings the state back to the code space, the last measurement round is assumed to be free of readout errors.
\roundd{}{This is a common way of eliminating a source of error that is not compounded when the quantum code is used in a fault-tolerant quantum computation.}

\begin{figure}[!ht]
\begin{subfigure}[b]{0.23\textwidth}
    \centering
    \includegraphics[width=\textwidth]{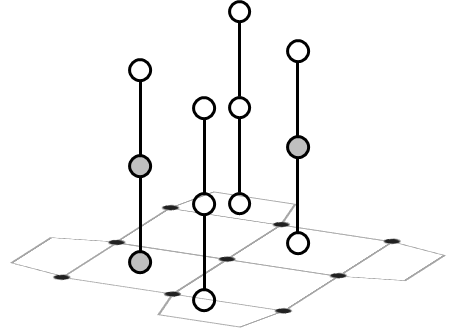}
    \caption{}
    \label{fig:mwpme}
\end{subfigure}
    \hfill
\begin{subfigure}[b]{0.23\textwidth}
    \centering
    \includegraphics[width=\textwidth]{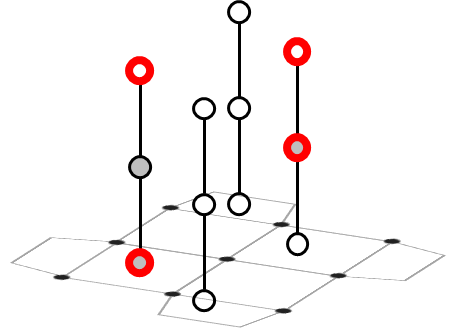}
    \caption{}
    \label{fig:mwpmf}
\end{subfigure}
\begin{subfigure}[b]{0.23\textwidth}
    \centering
    \includegraphics[width=\textwidth]{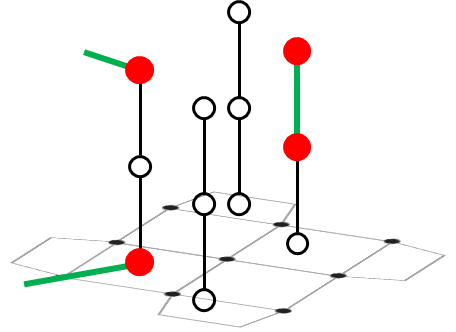}
    \caption{}
    \label{fig:mwpmg}
\end{subfigure}
    \hfill
\begin{subfigure}[b]{0.23\textwidth}
    \centering
    \includegraphics[width=\textwidth]{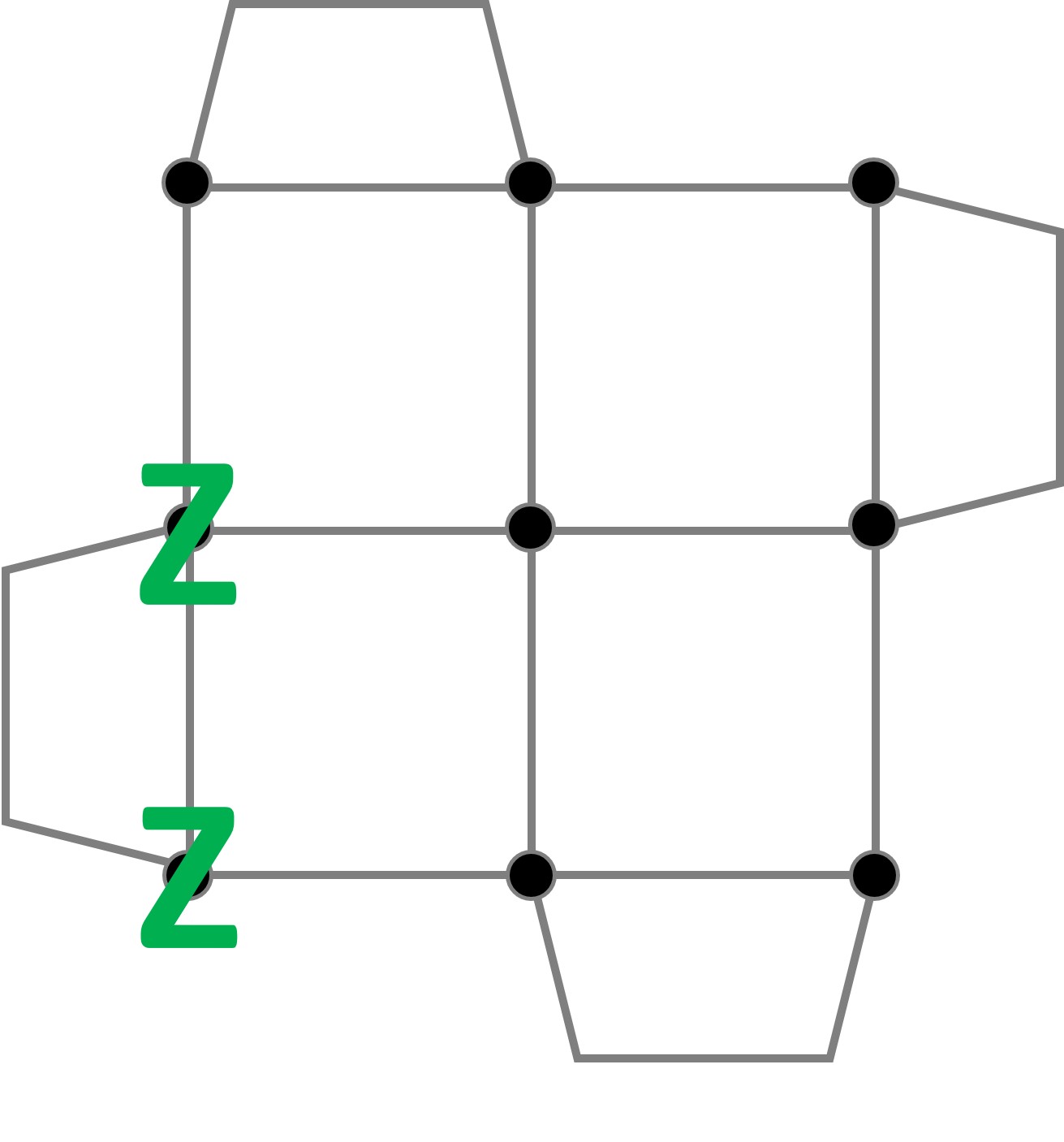}
    \caption{}
    \label{fig:mwpmh}
\end{subfigure}
\caption{Specific example of MWPM decoding method on a code patch with $d=3$. Stabilizer measurement outcomes are represented on a 3D grid, $\pm 1$ outcomes as white/grey circles. Vertices, where the measured value differs from the previous round, are marked with red circles. The minimum weight set of edges, which perfectly connects the marked vertices denoted by green color. This set of edges force the correction including 2 $\hat{Z}$ operators.}
\label{fig:incoherent}
\end{figure}

The final state, after the error correction operator has been applied, cf.~Eq.~\eqref{eq:final_state_coherent_readout}, 
reads
\begin{align} \label{eq:final state}
    \ket{\Phi_{\underline{s},\underline{s}'}}=e^{i \theta_L(\underline{s},\underline{s}') \hat{Z}^L}\ket{\psi_L}.
\end{align}
The logical rotation angle $\theta_L(\underline{s},\underline{s}')$ depends on perfect and noisy 3D syndrome too, as 
\begin{equation} \label{eq:theta_L} 
\begin{aligned}
    \theta_L  =\theta^* \qquad &\leftarrow \text{ }\hat{C}_{\underline{s}'}\hat{C}_{s_d}\ket{\psi_L}=\ket{\psi_L}; \\
    \theta_L =\theta^*+\dfrac{\pi}{2} \text{ } &\leftarrow \text{ }\hat{C}_{\underline{s}'}\hat{C}_{s_d}\ket{\psi_L}=\hat{Z}^L\ket{\psi_L}.
\end{aligned}
\end{equation}
Here the property that the operator $\hat{C}_{\underline{s}'}\hat{C}_{s_d}$ acts like a logical Z-operator or an identity is guaranteed by the constraint of perfect measurements in the last round of error correction.

The average diamond-norm distance and maximum infidelity can be expressed as:
\begin{align} \label{eq:dia_fid}
    P_L^d=2\sum_{\underline{s},\underline{s}'}P(\underline{s})P(\underline{s}\rightarrow \underline{s}')|\sin(\theta_L(\underline{s},\underline{s}'))|; \\
    P_L^i=\sum_{\underline{s},\underline{s}'}P(\underline{s})P(\underline{s}\rightarrow \underline{s}')\sin^2(\theta_L(\underline{s},\underline{s}')).
\end{align}

\section{Fermionic Linear Optics Simulation} 
\label{sec:Simulation_method}

We simulate quantum error correction by sampling the random outcomes of the syndrome measurements, and the final state of the logical qubit after the error correction.    
This can be summarized in the following steps:
\begin{enumerate}
    \item Generate a sample of the 3D syndrome $\underline{s}$ - a sequence of $d$ syndrome measurement rounds -  from the probability distribution $P(\underline{s})$.
    \item Calculate the corresponding rotation angle of the logical qubit, $\theta^*(\underline{s})$.
    \item Generate readout errors, i.e., noisy syndrome $\underline{s}'$ from probability distribution $P(\underline{s}\rightarrow \underline{s}')$.
    \item Calculate whether the rotation angle of the logical qubit is changed as a result of the readout errors, i.e., whether  $\theta_L=\theta^*+\pi/2$ or $\theta_L = \theta^*$, according to Eq.~\eqref{eq:theta_L}
\end{enumerate}
Steps 3. and 4. here are straightforward, they involve changing measured values of stabilizers with readout error probability $q$, and decoding using MWPM. Steps 1. and 2., however, can only be computed efficiently using Fermionic Linear Optics tools, and subtle tricks, as recently introduced by Bravyi et al \cite{bravyi2018correcting}. 
We only introduce some of the main concepts of the method of Bravyi et al. here, and give a brief summary in an Appendix; see \cite{bravyi2018correcting, benjamin_florian} for more details. We describe how the method can be extended to the sampling of repeated syndrome measurements.  

\subsection{Defining Majoranas for the qubits} \label{sec:FLO}

%

To make use of the tools of fermionic linear optics, we introduce a four-dimensional Hilbert space for each qubit, and four Majorana operators (Majoranas) acting in this Hilbert space. The Majoranas for the $m$-th qubit are denoted by $\hat{c}_j^{(m)}$, with $j=1,2,3,4$. 
They are similar to fermionic operators, in that different Majoranas anticommute; however, all Majoranas square to the identity, $\hat{c}_j^{(m)} \hat{c}_l^{(m')} + \hat{c}_l^{(m')} \hat{c}_j^{(m)} = 2 \delta_{jl} \delta_{m'm}$. 
Using the so-called C4 code \cite{bravyi2018correcting, kitaev2006anyons}, the Pauli operators acting on the qubit are represented using Majoranas as  
\begin{align} \label{eq:encoded pauli}
    \hat{X}_m &=i\hat{c}_1^{(m)} \hat{c}_2^{(m)}; \quad \hat{Z}_m =i\hat{c}_2^{(m)}\hat{c}_3^{(m)}; \nonumber \\ 
    \quad \hat{Y}_m &=i\hat{c}_3^{(m)}\hat{c}_1^{(m)}.
\end{align}
These operators fulfil the commutation relations expected of the Pauli operators. 

The Majoranas require a Hilbert space that is larger than that of the qubit itself. In fact, since above the Pauli operators were represented by products of two Majoranas, all states of the qubit are represented in so-called fixed-parity subspaces. 
We work on the subspace defined as $+1$ eigenspace of the C4 stabilizer,
\begin{align} \label{eq:c4 stab}
    \hat{S}^{(m)} =-\hat{c}_1^{(m)}\hat{c}_2^{(m)}\hat{c}_3^{(m)}\hat{c}_4^{(m)}.
\end{align}

The advantage of introducing Majoranas is that initialization of the code, coherent errors, and sampling the measurement statistics of the stabilizers can all be mapped to \roundd{free time evolution of}{the time evolution of a noninteracting fermionic system}. 

The main idea of the \roundd{FLO}{fermionic linear optics} approach is that we can work with the covariance matrix of the Majorana operators. 
Therefore instead of simulating the state vector of $d^2$ qubits, with $2^{d^2}$ elements, it is enough to keep track of the covariance matrix with $(2d)^4$ elements. 
With proper transformations of the covariance matrix, \roundd{which forced by}{corresponding to} free fermionic time evolution and measurement of Majorana pairs, we are able to sample $\theta^*(\underline{s})$ from the distribution $P(\underline{s})$ in $\mathcal{O}(d^4)$ time.   

We have extended the original simulation method for coherent errors \cite{bravyi2018correcting}, to the case of simultaneous coherent and readout errors. The key observation is that multiple rounds of coherent errors and stabilizer measurements can be decomposed into single rounds of inhomogeneous coherent errors, \roundd{(when}{, i.e., where} the physical rotation angle $\theta$ can be different for each physical qubit\roundd{)}{}. 

Starting from Eq.~\eqref{eq:final_state_start}, we are able to write it a slightly different way by inserting identities in the form $\hat{C}_{s_j}\hat{C}_{s_j}$,
\begin{equation}
\begin{aligned}
    \ket{\Phi_{\underline{s},\underline{s}'}}=\dfrac{1}{\sqrt{P(\underline{s})}}\hat{C}_{\underline{s}'}\hat{C}_{s_d}\hat{C}_{s_d}\hat{\Pi}_{s_d}\hat{U}\hat{C}_{s_{d-1}} \\
    \hat{C}_{s_{d-1}}\hat{\Pi}_{s_{d-1}}\hat{U}...\hat{C}_{s_1}\hat{C}_{s_1}\hat{\Pi}_{s_1}\hat{U}\ket{\psi_L}.
\end{aligned}
\end{equation}
Furthermore each round can be written in the form of Eq.~\eqref{eq:final_coherent_start}, with $\hat U$ replaced by inhomogeneous error operators $\hat{U}_j=\hat{U}\hat{C}_{s_{j-1}}$, and the normalization factor by $1/\sqrt{P(s_j)}$. Based on Eq.~\eqref{eq:final_state_coherent}, we can express each round as a rotation about the Z axis,
\begin{align} \label{eq:inhom_rounds}
    \dfrac{1}{\sqrt{P(s_j)}}\hat{C}_{s_j}\hat{\Pi}_{s_j}
    \!\underbrace{\hat{U}\hat{C}_{s_{j-1}}}_{\hat{U}_j}\ket{\psi_L}\! =e^{i  \theta_j^L\hat{Z}^L}\ket{\psi_L},   
\end{align}
where the logical rotation angle $\theta^L_j$ depends on all the previously measured syndromes ($s_1,s_2,..,s_j$).

Finally one can write the final state for perfect syndrome $\underline{s}$, and noisy syndrome $\underline{s}'$ as
\begin{align}
\ket{\Phi_{\underline{s},\underline{s}'}}=\hat{C}_{\underline{s}'}\hat{C}_{s_d}e^{i\theta ^*(\underline{s})\hat{Z}^L}\ket{\psi_L},
\end{align}
where the rotation angle $\theta ^*(\underline{s})$ can be calculated from sampling single rounds of error correction with perfect syndromes and inhomogeneous coherent errors,
\begin{align}
    \theta ^*(\underline{s})=\sum_{j=1}^d\theta_j^L(s_1,s_2,...,s_j).
\end{align}

\section{Numerical results} \label{sec:Results}

We used the fermionic linear optics method to simulate the surface code under coherent and readout errors, for code sizes up to $d=19$. \roundd{We}{As detailed below, we} sampled the logical rotation angle distribution, from which we computed -- for the most susceptible initial states -- 
\roundd{both the expecation value of the infidelity and the diamond norm distance to the identity channel}
{the expectation value of the infidelity, which we call logical error rate}.
As code sizes were scaled up, we found threshold behaviour 
\roundd{with both types of error measures (although somewhat less convincingly with the diamond norm)}{}. 
In case the rates of coherent and readout errors were equal, $p=q$, we found that the threshold is close to the corresponding threshold of random Pauli Z + readout errors.
Our results here are similar to those with perfect measurements by Bravyi et al\cite{bravyi2018correcting}.

We also investigated how, below the threshold, the logical error rates compare to those of the random Pauli Z + readout errors, and how the residual coherence in the logical error decreases as the code size is scaled up.    
\roundd{We}{As detailed below, we} again find similar results to those with perfect measurements by Bravyi et al.\cite{bravyi2018correcting}.
Varying the rates of the two error processes independently, we mapped out the threshold on the ($p,q$) plane, and found that coherent errors are more critical than readout errors: to achieve scalable error correction it is easier to compensate a high value of readout errors (at or above 10\%) by reducing the rate of coherent errors than vice versa.

\subsection{Threshold with equal coherent and readout errors}

\begin{figure}
    \includegraphics[width= \columnwidth]{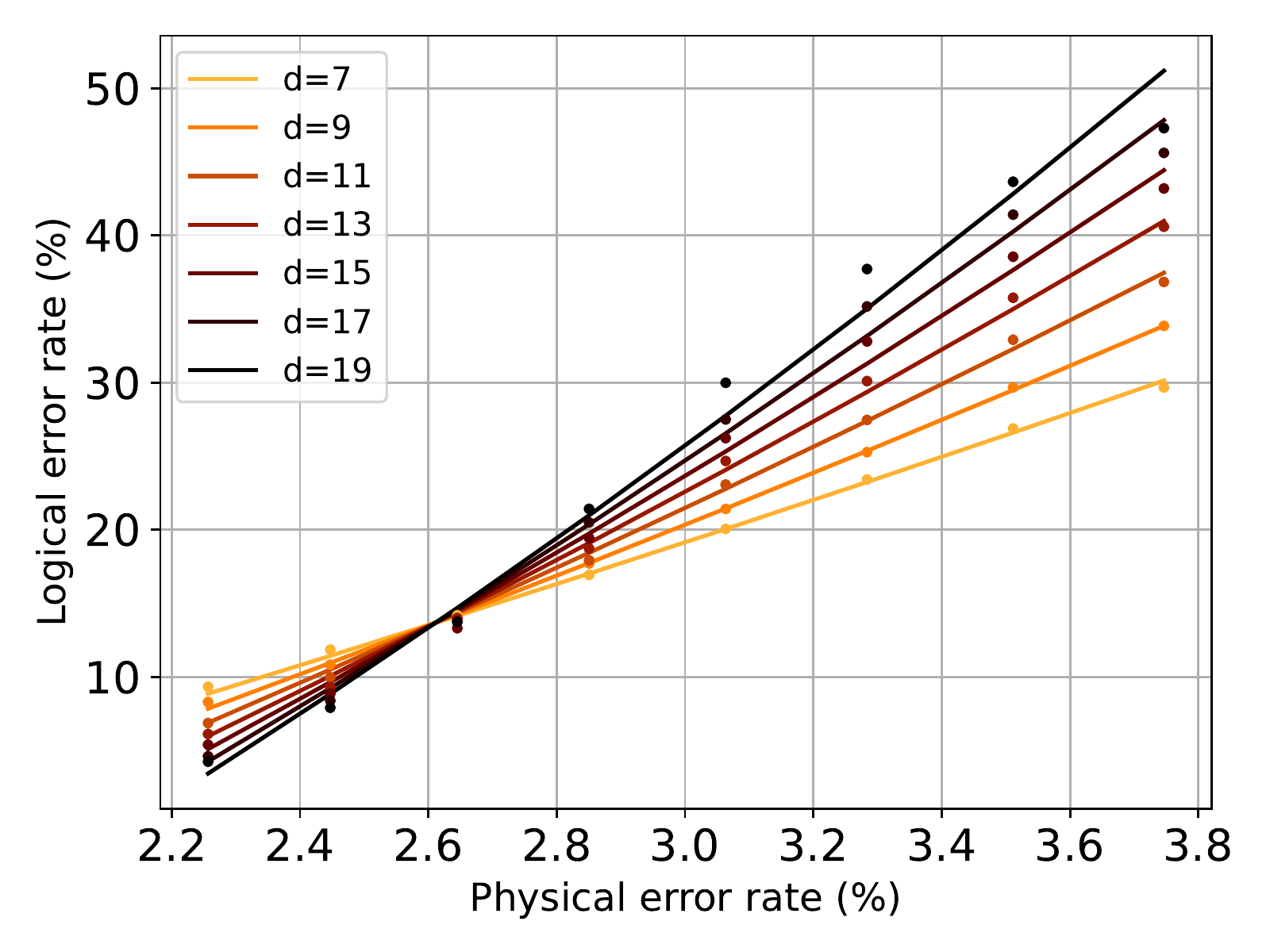}
    \caption{Error threshold as the code size is scaled up with coherent and readout error rates equal ($p=q$ is the "physical error rate"). 
    \roundd{The numerically obtained diamond-norm distance (a), and the maximum infidelity (logical error rate) (b)}
    {Logical error rates for different code distances}
    show threshold behaviour as functions of the physical error rate. \roundd{}{For the fitted curves we used a finite size scaling ansatz\cite{wang2003confinement}, based on the statistical mechanical mapping of the surface code.} Every point results from $5000\times d\times 100$ rounds of simulation.}
    \label{fig:thresholds}
\end{figure}

We ran extensive simulations to estimate the error threshold when the readout error rate $q$ is set equal to the coherent error rate $p$. 
For every odd value of code distance $d$, up to $d=19$, to obtain the numerical distribution of logical rotation angles, we sampled the noiseless 3D syndrome measurements $5000$ times (5000 $d$ rounds), and then sampled 100 noisy syndromes from each of these. 
In Fig.~\ref{fig:thresholds}, we show the resulting \roundd{average diamond norm distance and the}{} 
average logical error rate, calculated via Eqs.~\eqref{eq:dia_fid}.  
\roundd{Using both measures, we}{We} observe that for errors below a threshold, scaling up the code size decreases the logical errors, while above the threshold, scaling up only makes things worse by increasing the logical errors. 

To obtain a precise value of the threshold, we fitted the numerical values using a finite size scaling ansatz\cite{wang2003confinement}, based on mapping of the surface code to statistical physics models\cite{kitaev_preskill2002}.
Although the ansatz is strictly expected to work for random Pauli+readout errors, it also fits our numerics (coherent+readout errors). \roundd{albeit less convincingly for the case of the diamond norm.}{} 
The threshold value is
\begin{equation} \label{eq:thresholds}
    p_{th}^i=2.62\%\pm 0.02\%. 
\end{equation}
\roundd{The threshold via the diamond norm is significantly higher than that via the infidelity, but both are}{This is} relatively close to the threshold of random Pauli + readout errors\cite{wang2003confinement}, which for the toric code is $p_{th}=2.93\%\pm0.02\%$.
\roundd{}{Using the diamond-norm distance, our numerics paints a somewhat different picture, as discussed in Appendix~\ref{apx:diamond_norm}.}

\subsection{Sub-threshold comparison with random Pauli+readout errors}

\begin{figure}
\centering
\includegraphics[width=\columnwidth]{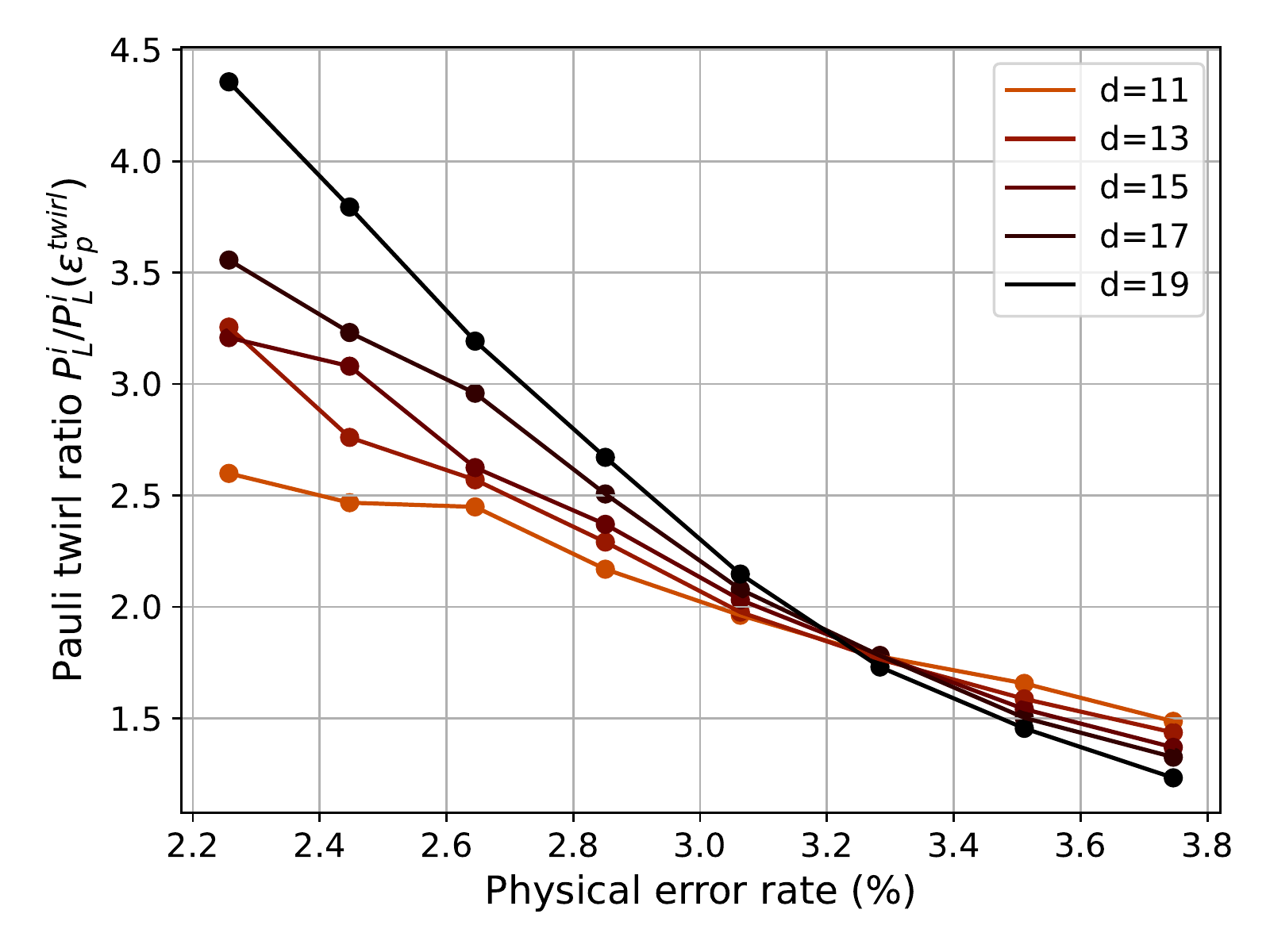}
\caption{
    Pauli twirl ratio (for maximum infidelity) as functions of physical error rate (equal to readout error rate), for different code distances. Results obtained from the same data as in Fig.~\ref{fig:thresholds} and Monte Carlo simulations of incoherent noise.
    }
    \label{fig:Pauli_twirl_ratios}
\end{figure}

 We next compare the performance of the surface code under coherent+readout errors with its performance under random Pauli+readout errors, when error rates are below threshold. We use the same parameter \roundd{}
 {$p=\sin^2(\theta)$} 
 for the two channels, with the random Pauli channel being the Pauli twirled version of the coherent error channel,    
\begin{align}
    \varepsilon^{twirl}_{\theta}(\hat{\rho})=\cos^2(\theta)\hat{\rho}+\sin^2(\theta)\hat{Z}\hat{\rho}\hat{Z}.  
\end{align}
The \emph{Pauli twirl ratio} is the ratio of logical error rate in case of coherent errors+readout errors, and the logical error rate when the coherent errors are replaced by their Pauli twirled version \roundd{}{at the physical level}. 
\roundd{We denote this latter quantity with $P_L^d(\varepsilon_p^{twirl})$ and $P_L^i(\varepsilon_p^{twirl})$ when the diamond norm and the infidelity are used, respectively.}
{Therefore Pauli twirl ratio can be written as:
\begin{align}
    \dfrac{P^i(\varepsilon_{\theta})}{P^i(\varepsilon_{\theta}^{twirl})}; \qquad  \dfrac{P^d(\varepsilon_{\theta})}{P^d(\varepsilon_{\theta}^{twirl})},
\end{align}
for the maximum infidelity and for the diamond-norm distance.}

The numerically obtained values of the twirl ratio, shown in Fig.~\ref{fig:Pauli_twirl_ratios}, indicate that  below the threshold, coherent errors + readout errors lead to higher logical error rates than random Pauli + readout errors (high Pauli twirl ratio). 
Moreover, this difference grows as we scale up the size of the code. 
\roundd{}{Note that in the limit of vanishing error rates, $p\to 0$, 
we expect the Pauli twirl ratio to reach a finite, code distance dependent value. The reason is that we expect the logical error rate for small $p$ to scale as 
$P_L \approx N_d  p^{(d+1)/2}$ for incoherent, and $P^i_L \approx N^2_d(\sin^2(\theta))^{(d+1)/2}$ for coherent errors. 
Here $N_d$ is the number of the shortest errors strings that cause a logical error, and it should be roughly the coherence ratio in the $p \rightarrow 0$ limit.
 Numerical investigation of the $p\to 0$ limit is, however, expensive due to the small error rates. }

Interestingly, there is a threshold-like behavior of the Pauli twirl ratio, which is independent of code distance for  $p \approx 3.3\%$, and decreases with code distance for $p$ above. However, the value $3.3\%$ is not the threshold of the code.

One of the key findings of Bravyi et al \cite{bravyi2018correcting} is that quantum error correction "washes out" coherence in the logical level in the coherent errors (without readout errors). As the code distance increases, the distribution of logical rotation angles becomes more and more highly peaked around 0 and $\pi/2$, thus, the logical noise is better and better approximated by random Pauli process\roundd{}{es}. This property of quantum error correction has also been studied analytically\cite{beale2018quantum, iverson_preskill}.
A practical quantity to study this effect is the \emph{coherence ratio}\cite{bravyi2018correcting}, defined as
\begin{align}
    \dfrac{P^d_L}{P_L^{d,twirl}}=\dfrac{\sum_{\underline{s},\underline{s}'}p(\underline{s})p(\underline{s}\rightarrow\underline{s}')|\sin(\theta_L)|}{\sum_{\underline{s},\underline{s}'}p(\underline{s})p(\underline{s}\rightarrow\underline{s}')\sin^2(\theta_L)},
\end{align}
\roundd{}{where $P^{d,twirl}_L$ is just the diamond-norm distance for the twirled logical error channel (which is just the maximum infidelity up to a factor of 2).}
The coherence ratio is always greater than or equal to one, equality holds if $\theta_L$ only takes values $\{0,\pi/2\}$, i.e., if the logical noise is fully incoherent (probabilistic logical Z errors).

\begin{figure}
    \centering
    \includegraphics[width=\columnwidth]{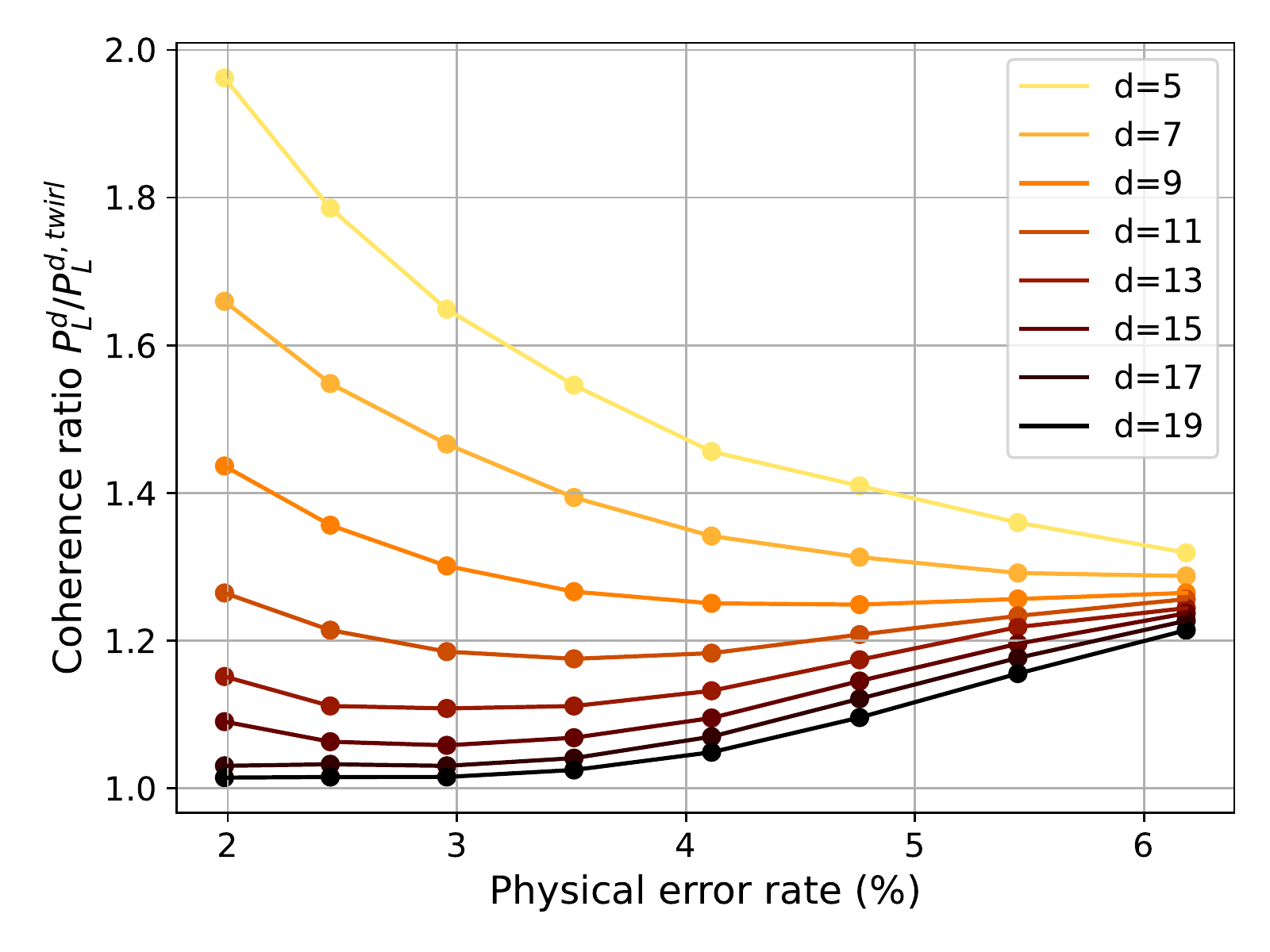}
    \caption{Coherence ratio as a function of error rate ($p=q$) with different code distances. Each point results from $5000\times d\times 100$ rounds of simulation.}
    \label{fig:coherence_ratio}
\end{figure}

One would expect that this "washing out" of coherence is, if anything, made even stronger by the readout errors. Our numerics confirms this intuition. In Fig.~\ref{fig:coherence_ratio} we show the coherence ratio of different code sizes with physical error rate and readout error rate set equal. We find that the coherence ratio decreases as the code is scaled up, even above the threshold. Around the threshold, $p=q \approx 3\%$, practically all coherence is lost on the logical level, for code sizes $d=17$ and above. Interestingly though, the coherence ratio appears to increase as the physical error rate is decreased from the threshold. Without readout errors\cite{bravyi2018correcting}, all of these qualitative trends are there, but the coherence ratio is $1.1$ at the threshold even for a code size $d=37$. 

\subsection{Independent coherent and readout errors}

We have also investigated how varying the rate $p$ of coherent errors and $q$ of readout errors independently affects the threshold of the surface code. 
For many pairs of $p$ and $q$ we numerically ascertained whether scaling up the surface code decreases logical error rates (scalable QEC) or it increases them (unscalable QEC). The threshold should be in between these regions. 
We could not determine the threshold values more precisely, since the fitting ansatz we used for $p=q$ turned out to be a poor fit in many of the cases with asymmetric noise.

\begin{figure}
    \includegraphics[width= \columnwidth]{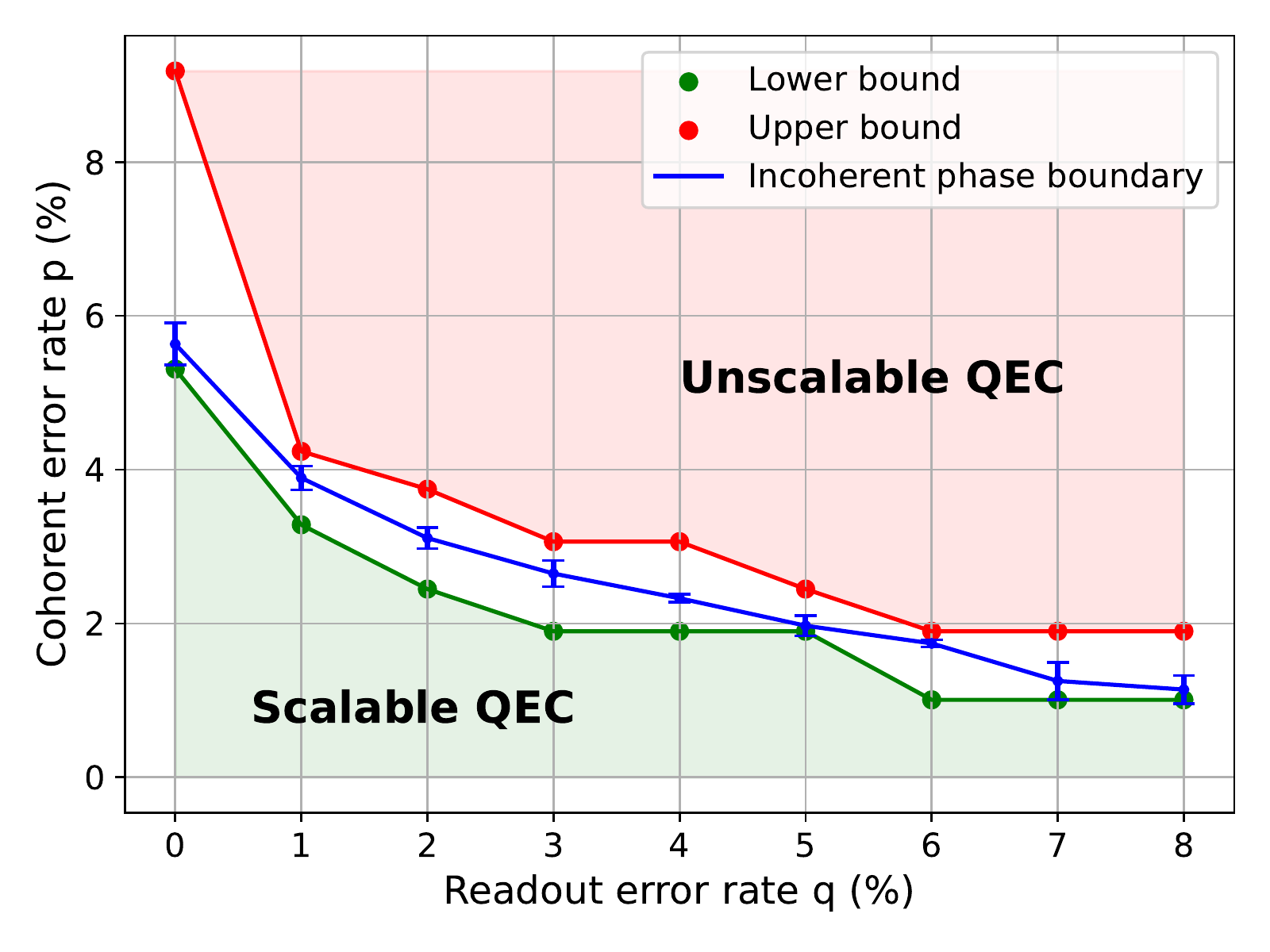}
    \caption{Threshold lines on the ($p,q$) plane for maximum infidelity (logical error rate). For a fix $q$ value we run the simulation for $16$ different $\theta$ values, and determined a lower and an upper bound for thresholds in each cases. The lower bound is the last point where the maximum infidelity is decreasing as the code distance is increasing. The upper bound however is the first point where the logical error rate is increasing as the distance is increasing. We numerically investigated the incoherent case via Monte Carlo simulations, and determined the threshold values for asymmetric $p,q$ values with the fitting ansatz\cite{wang2003confinement}.}
    \label{fig:threshold_map}
\end{figure}

Our results, shown as a 2D map in Fig.~\ref{fig:threshold_map}, show that the surface code is more sensitive to coherent errors than to readout errors. If the coherent error rate is on the percent level, the surface code is quite robust against readout errors, scalable even with relatively high $q \approx 7\%$. However, if the readout error rate is on the percent level, the surface code still requires the coherent error rate to be below $3\%$.   

\section{Discussion and outlook}
\label{sect:Discussion}

We investigated numerically how well the surface code works as a quantum memory when there are coherent errors on the physical qubits as well as readout errors (phenomenological readout error model). We focused on a restricted class of coherent errors, namely, unitary phase rotations, $e^{i\theta \hat{Z}_j}$. This allowed us to use the theoretical tools of Fermionic Linear Optics, as applied to the surface code by Bravyi et al\cite{bravyi2018correcting}. We extended that work by including readout errors as well, on a phenomenological level (perfect measurements, noisy recording of measurement results).
\roundd{}
{The python source code for our numerical work is available on GitHub\cite{code}.}

Our results show that the findings of Bravyi et al\cite{bravyi2018correcting} on the effects of coherent errors on the surface code mostly carry over to when readout errors also occur. Namely, 
the surface code with coherent+readout errors has a threshold, which is close to that of the corresponding Pauli twirled error channel (random Pauli+readout errors). However, for error rates below the threshold, its logical error rates are significantly higher than that of the Pauli twirled error channel. Scaling up the code size, coherence is washed out from the logical error. Moreover, we found that having a low value of the coherent errors is more important than a low value of readout errors (high readout error rates can be compensated by low coherent error rates, but less so vice versa). 

A point that is worth further investigation is the differences in our results when using the diamond norm or the fidelity as quantitative measures of the reliability of quantum memory. \roundd{Although they gave qualitatively similar results, they were quantitatively different: e.g., the value of the threshold was significantly higher (by 0.5\%) when using the diamond norm.}
{As an example, these two measures gave quite different results regarding the threshold of coherent+readout errors: we observed a clear threshold using the fidelity, but less clear behavior using the diamond norm (Appendix \ref{apx:diamond_norm}).}

It would also be interesting to consider broader classes of coherent errors. A next step would be to consider coherent error parameters $\theta$ that are not constant, but vary from qubit to qubit or even fluctuate (this latter case modeling the combination of coherent and incoherent Z errors). A numerically more challenging question is how our results would be changed if even the axis of coherent rotation varied from qubit to qubit (not $Z$ for all qubits as in our work) --  unfortunately here the tools Bravyi et al.\cite{bravyi2018correcting} do not apply. 
Even more challenging is to bring the error model closer to experimental reality, by modeling coherent errors on the circuit level. 
\roundd{For this case recent theoretical work }
{A step in this direction was taken recently, with a coherent error model that includes ancilla qubits, but uses a somewhat unrealistic multiqubit gate}
\cite{zhao2023lattice}
\roundd{, using}
{. For this error model, }
a mapping to 
\roundd{}{a} 
three dimensional lattice gauge theory seems to suggest that when combined with incoherent errors, coherent errors ruin the threshold: even with arbitrarily small error rates, scaling the code size up beyond a certain size will increase noise the logical level. 
\roundd{}{A related point is that effects of coherent errors can be mitigated in the surface code using extra ancilla qubits to realize code concatenation\cite{hu2021mitigating, ouyang2021avoiding}, or in variants of  Shor's code by other means\cite{debroy2021optimizing}. }

\section*{Acknowledgements}

This research was supported by the Ministry of Culture
and Innovation and the National Research, Development
and Innovation Office (NKFIH) within the Quantum Information National Laboratory of Hungary (Grant No.
2022-2.1.1-NL-2022-00004), and by the NKIFH within
the OTKA Grant FK 132146.
This research has been supported by the Horizon Europe research and innovation programme of the European Union through the IGNITE project\roundd{}{
and by the HORIZON-CL4-2022-QUANTUM01-SGA project 101113946
OpenSuperQPlus100 of the EU Flagship on
Quantum Technologies}. 
\roundd{}{We also acknowledge support by the ÚNKP-22-2-II-BME-10 New National Excellence Program of the Ministry for Culture and Innovation from the source of the National Research, Developement and Innovation Fund.}

\bibliographystyle{quantum}
\bibliography{main}

\appendix

\section{Results for diamond-norm distance}
\label{apx:diamond_norm}

We have omitted some of the results for diamond-norm distance from the main part of our paper, because \roundd{we found them very similar to the results for maximum infidelity.}{} we think that maximum infidelity is a more convenient measure to use, but for the sake of completeness here we show the diamond-norm distance versions of Fig.~\ref{fig:thresholds}, Fig.~\ref{fig:Pauli_twirl_ratios}, and Fig.~\ref{fig:threshold_map}.

\begin{figure} [!h]
    \centering
    \includegraphics[width=\columnwidth]{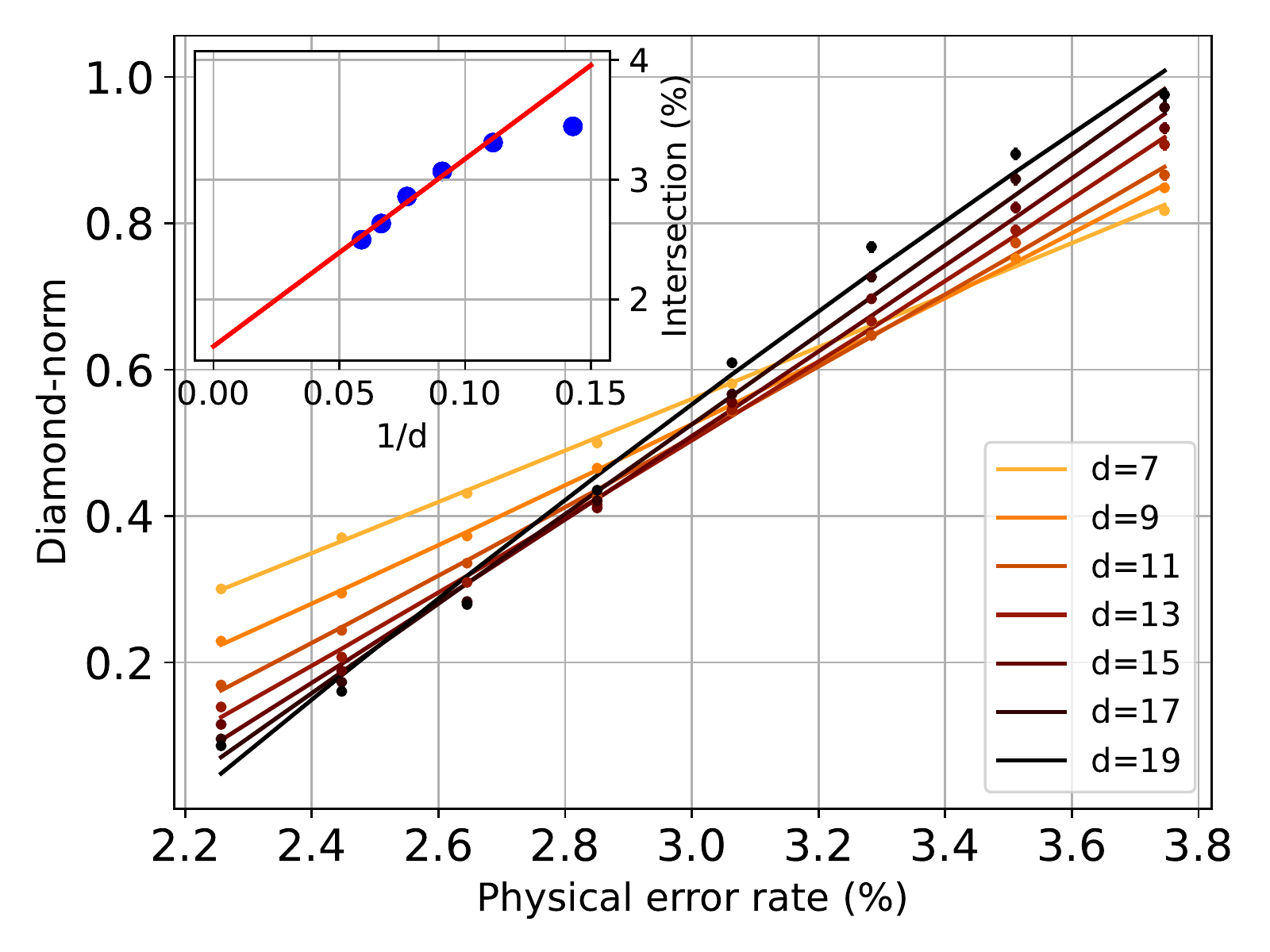}
    \caption{\roundd{}{Diamond-norm distance as the function of physical error rate for several code distances. We fitted second order polynomials for each code distance, and plotted the intersections of these polynomials as the function of the inverse code distance. (At 1/d we placed the intersection of polynomials that correspond to code distances d and d+2.) We fitted a linear function to the first five data points, and read the $1/d = 0$ value as the threshold. Results obtained from the same data as in Fig.~\ref{fig:thresholds}}
    \label{fig:dia_threshold}}
\end{figure}

\begin{figure} [!ht]
    \centering
    \includegraphics[width=\columnwidth]{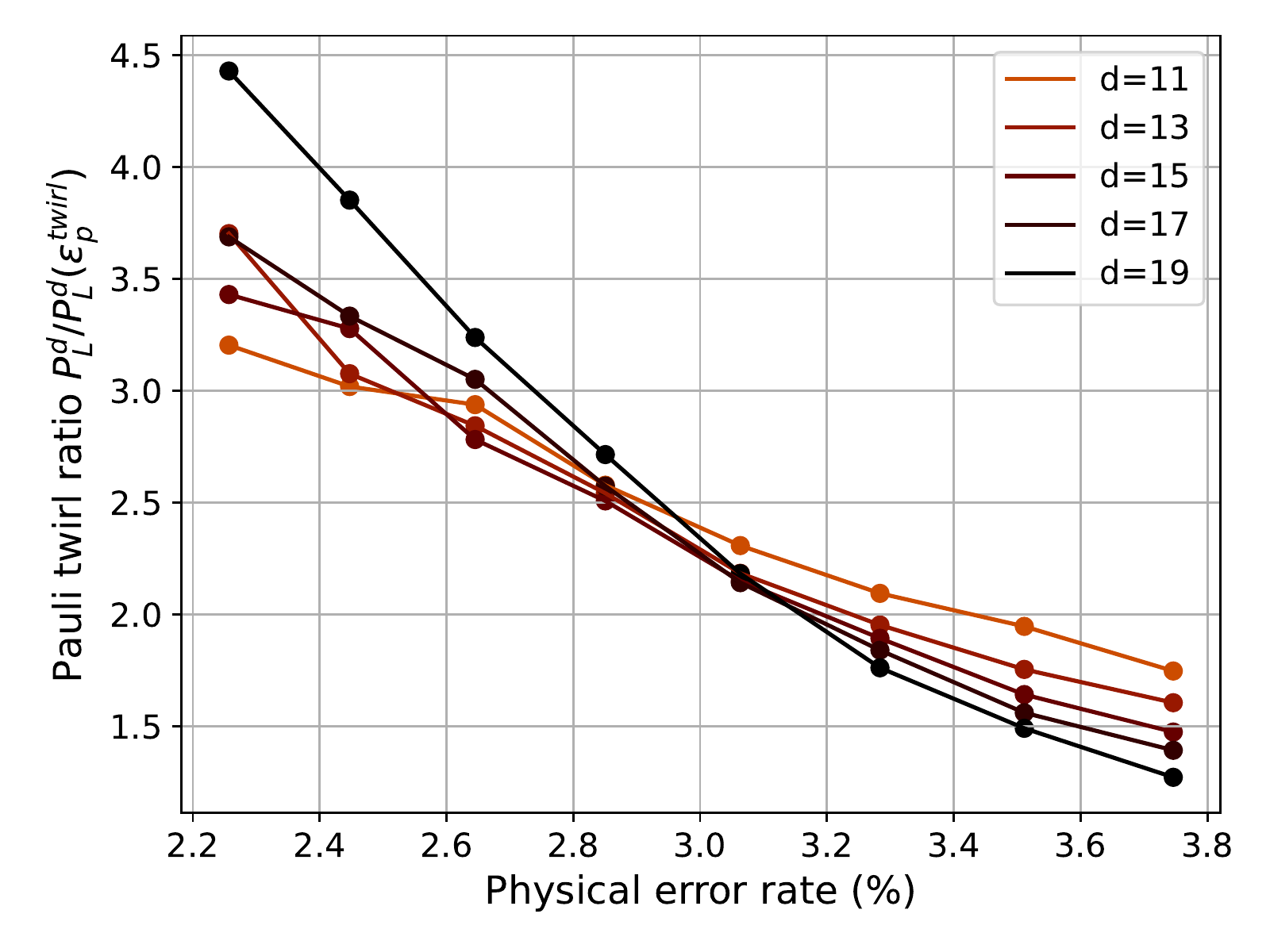}
    \caption{Pauli twirl ratio (for diamond-norm distances) as the function of physical error rate. Results obtained from the same data as in Fig.~\ref{fig:Pauli_twirl_ratios}}
    \label{fig:dia_twirl_ratio}
\end{figure}

\begin{figure} [!ht]
    \centering
    \includegraphics[width=\columnwidth]{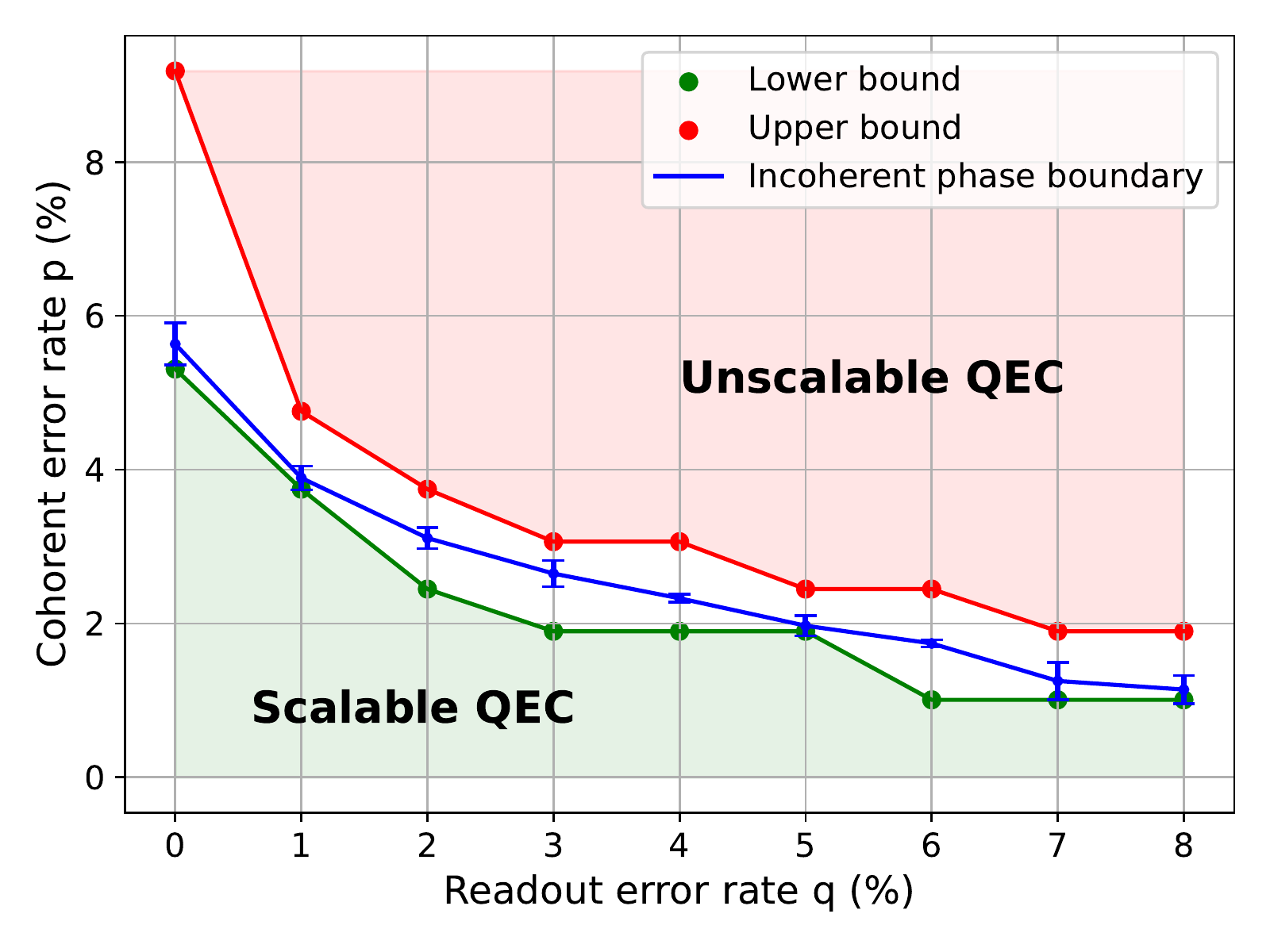}
    \caption{Threshold lines on the (p,q) plane for diamond-norm distance. Results obtained from the same data as in Fig.~\ref{fig:threshold_map}}
    \label{fig:dia_threshold_map}
\end{figure}

\roundd{}
{As far as the error threshold is concerned, we did not find the same convincing numerical evidence using the diamond-norm distance as we did with the infidelity. As shown in Fig.~\ref{fig:dia_threshold}, the plots of the diamond-norm distance for different code sizes (quadratic curves fitted individually) do not intersect in the same point. Rather, the physical error rate where the curve of size $d$ intersects that of the curve for size $d+2$ is smaller for larger code sizes. 
We try to estimate the threshold, i.e., a limit of the intersection points in the $d\to \infty$ limit, by plotting in the inset of Fig.~\ref{fig:dia_threshold} the physical error rates of these intersection points as the function of the inverse code size $1/d$. The data suggests a a threshold around $p=1.6\%$. 

For the sake of completeness, we also include here plots of the Pauli twirl ratio, evaluated using the diamond-norm distance, Fig.~\ref{fig:dia_twirl_ratio}, and a 2D map of how the finite-size threshold values depend on the coherent error rate $p$ and readout error rate $q$, Fig.~\ref{fig:dia_threshold_map}. 
These both show qualitatively and also quantitatively similar behavior to their infidelity-based counterparts,  
Figs.~\ref{fig:Pauli_twirl_ratios} and \ref{fig:threshold_map}, respectively.}

\section{Sampling Coherent Errors in the Surface Code with Fermionic Linear Optics}

Here we summarize the technical details of our simulations, which were based on the original proposal by Bravyi et al \cite{bravyi2018correcting}. Previously we have shown Eq.~\eqref{eq:inhom_rounds} that the simulation of coherent+readout errors can be done by simulating error correction rounds with inhomogeneous coherent errors. Therefore here we restrict our description to single error correction rounds with perfect readout. 

First we introduce an exact mapping from the surface code to a fermionic system which can be described by Majorana operators. We show the \roundd{FLO}{fermionic linear optics} algorithms for transforming the covariance matrix of the fermionic system. Then we describe the exact steps for sampling $\theta_L(s)$ from the probability distribution $P(s)$. 

The aim of this Appendix is to help to understand the details of the \roundd{FLO}{fermionic linear optics} algorithm\cite{bravyi2018correcting}. For further details our code is also available on GitHub\cite{code}.

\subsection{Encoding qubits into fermionic systems}

As we mentioned in Sec.(\ref{sec:FLO}) physical qubits of the surface code can be represented as C4 codes.
However this mapping is not just a mathematical transformation, but a physically well-motivated thing.

The basic idea is to encode the physical qubits into double quantum dots with 2 fermionic modes described by creation and annihilation operators $\hat{a}^{(m)}_1,\hat{a}^{\dagger (m)}_1,\hat{a}^{(m)}_2,\hat{a}^{\dagger (m)}_2$ for the $m$-th qubit. With these fermionic operators in hand we can define the usual qubit basis states from the fermionic vacuum for one C4 code,
\begin{equation}
    \ket{0}=\hat{a}^{\dagger}_2\ket{\emptyset}; \quad \ket{1} = \hat{a}^{\dagger}_1\ket{\emptyset}.
\end{equation}
We can introduce Majorana operators in the following way:
\begin{align}
    \hat{c}_1=i(\hat{a}_2-\hat{a}^{\dagger}_2); \qquad \hat{c}_2=\hat{a}_1+\hat{a}^{\dagger}_1; \\
    \hat{c}_3=i(\hat{a}_1-\hat{a}^{\dagger}_1); \qquad
    \hat{c}_4=\hat{a}_2+\hat{a}^{\dagger}_2.
\end{align}
As we define these Majorana operators for each qubit we are able to express encoded Pauli operators $\hat{X}_m$, $\hat{Z}_m$, $\hat{Y}_m$ and C4 stabilizers $\hat{S}^{(m)}$, as we did in Eqs.~\eqref{eq:encoded pauli}\eqref{eq:c4 stab}. 

It is important to note that $\hat{S}^{(m)}\hat{X}_m$, $\hat{S}^{(m)}\hat{Z}_m$ and $\hat{S}^{(m)}\hat{Y}_m$ are also good Pauli operators, we will take advantage of this later.

\begin{figure}[!ht]
    \centering
    \includegraphics[width=.4\textwidth]{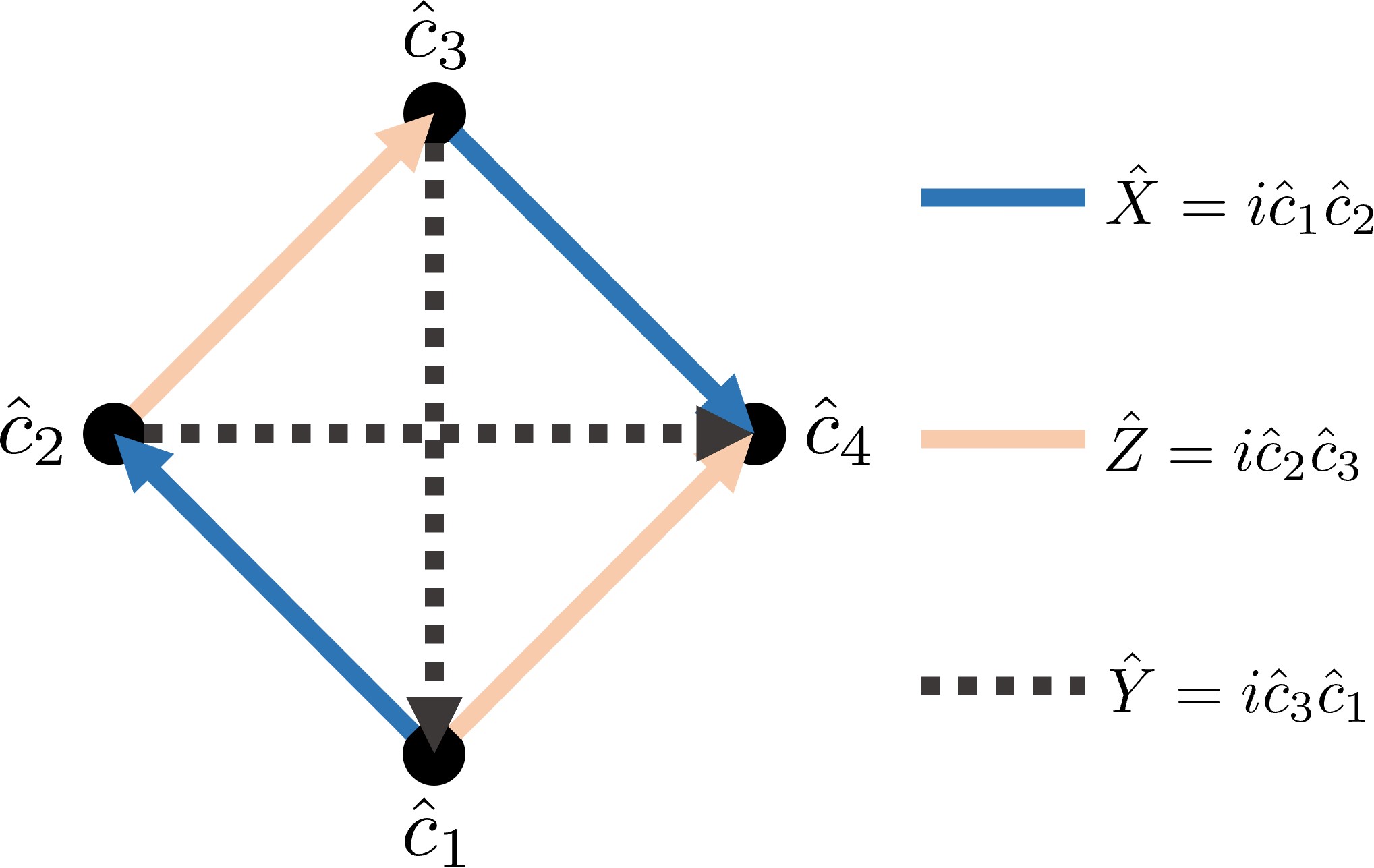}
    \caption{Visualization of one C4 code. Black dots represent Majorana fermions and arrows represent encoded Pauli operators (first Majorana operator is the tail second is the head). We have drawn the products of C4 stabilizer and Pauli operators too.}
    \label{fig:my_label}
\end{figure}

\subsection{Majorana representation of surface code}

\begin{figure}[!ht]
    \centering
    \includegraphics[width=.8\columnwidth]{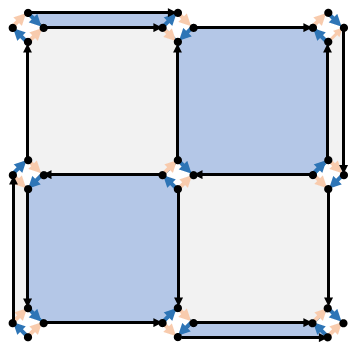}
    \caption{Basic building block of Majorana representation of the surface code. The orientation of link operators ensures Eq.~\eqref{eq:link operators orientation}. Bigger patches can be achieved by obvious duplication and translation of this small block.}
    \label{fig:majorana network}
\end{figure}

With the help of encoded C4 codes now we can build up the whole surface code, and identify the Majorana representation of stabilizers and logical operators.
First define some Majorana pairs, which connect Majorana fermions belong two different C4 codes, we call them \emph{link operators},
\begin{align}
    \hat{L}_e=i\hat{c}_p\hat{c}_q.
\end{align}
Here we get rid of the superscript index of the Majorana operators, and we use $p,q$ indices instead. These are at the ends of edge $e$.
Link operators assigned to every edge of the code patch, but the order of the Majorana operators is not well-defined (we have some freedom in the orientation of arrows, which represents link operators). The only restriction is that the product of link operators on the boundaries of faces have to be equal to the stabilizers on these specific faces,
\begin{align} \label{eq:link operators orientation}
    \prod_{e\in \partial f}\hat{L}_e=\hat{S}_f
\end{align}
We can fulfill this restriction by building up our code patch from basic building blocks, see Fig~\ref{fig:majorana network}. Note that every other C4 code is rotated by $90^{\circ}$, so only "proper" sides of C4 codes join to proper faces (X/Z operators to X/Z stabilizers).

\begin{figure}[!ht]
    \centering
    \includegraphics[width=\columnwidth]{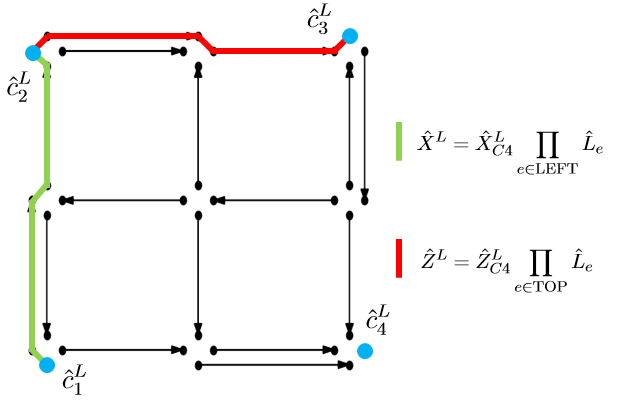}
    \caption{Logical operators and logical C4 code of the Majorana representation of the surface code.}
    \label{fig:logical majoranas}
\end{figure}

Next step is the expression of logical operators with the Majorana fermions of C4 codes. First we can observe that every Majorana fermion is part of a link operator, expect 4 corner Majoranas (see Fig. \ref{fig:logical majoranas}), from these Majoranas we can build up the so-called \emph{logical C4 code} with logical C4 Pauli operators analogous to Eq.~\eqref{eq:encoded pauli}:
\begin{align} \label{eq:logical C4}
    \hat{X}^L_{C4}=i\hat{c}^L_1\hat{c}^L_2; \quad \hat{Z}^L_{C4}=i\hat{c}^L_2\hat{c}^L_3.
\end{align}
Original logical operators of the surface code can be expressed with the help of these logical C4 operators and link operators based on the definition Eq.~\eqref{eq:logical operators}:
\begin{align}
    \hat{X}^L=\hat{X}^L_{C4}\prod_{e\in \text{LEFT}}\hat{L}_e \qquad \hat{Z}^L=\hat{Z}^L_{C4}\prod_{e \in \text{TOP}}\hat{L}_e.
\end{align}

One of the most important thing for the simulation is that we can express any logical state of the surface code with the help of Majorana operators. First define the state stabilized by all link operators,
\begin{align}
    \hat{L}_e\ket{\Phi_{link}}=\ket{\Phi_{link}} \quad \forall e. 
\end{align}
Note this $\ket{\Phi_{link}}$ state is not an eigenstate of C4 codes, and it does not determine the state of logical Majoranas. The logically encoded version of state $\ket{\psi}$ can be expressed as
\begin{align} \label{eq:Majorana_logical_state}
    \ket{\psi_L}=2^{(n-1)/2}\prod_{j=1}^n\dfrac{1}{2}(1+\hat{S}^{(j)})\ket{\Phi_{link}}\otimes\ket{\psi^L_{C4}},
\end{align}
where $\ket{\psi^L_{C4}}$ is the encoded version of $\ket{\psi}$ into the logical C4 code.

One can easily check that this state is the $+1$ eigenstate of all the surface code stabilizers and the logical operators of the surface code are acting just as encoded logical C4 operators on $\ket{\psi^L_{C4}}$.

\subsection{Fermionic Linear Optics}

Fermionic Linear Optics (FLO) \cite{FLO1,FLO2,bravyi2004lagrangian} simulations based on the fact that some specific fermionic states, consist of $N$ fermions, can be fully characterized by just a simple $2N\times2N$ matrix, these states are called \emph{pure Gaussian states}, and the matrix is the \emph{covariance matrix},
\begin{align}
    M_{jk}(\ket{\psi})=\bra{\psi}i\hat{c}_j\hat{c}_k\ket{\psi}-i\delta_{jk}.
\end{align}
Here Majorana operators ($\hat{c}_1,\hat{c}_2,...,\hat{c}_{2n}$) describe the fermionic system of $n$ fermions.
Covariance matrix can be written in every state, but in pure Gaussian states it will be orthogonal. (It can be an alternative definition, and it is also easy to check). The most important fact is that pure Gaussian states are fully characterized by their covariance matrix \cite{FLO2}, so simulation of pure Gaussian states can be realized by just tracking a matrix with $\mathcal{O}(n^2)$ elements.

\bigbreak

The limitation of \roundd{FLO}{fermionic linear optics} simulations is that we can only apply operations that bring the state from a pure Gaussian state to an other pure Gaussian state. During the simulation of coherent errors in the surface code 3 kind of enabled operations appear\cite{bravyi2018correcting}:
\begin{itemize}
    \item Initialization of states stabilized by Majorana pairs $i\hat{c}_p\hat{c}_q$.
    \item Application of operator $e^{\theta\hat{c}_p\hat{c}_q}$.
    \item Measurement of Majorana pair $i\hat{c}_p\hat{c}_q$.
\end{itemize}
For clarifying these operations we express the covariance matrices induced by them.

The covariance matrix of a pure Gaussian state $\ket{GS}$ stabilized by $2n$ ($p,q$) pairs,
\begin{align}
    i\hat{c}_p\hat{c}_q\ket{GS}=\ket{GS} \qquad \forall (p,q),
\end{align}
can be expressed as:
\begin{align}
    M_{jk}=\sum_{(p,q)}\delta_{jp}\delta_{kq}-\delta_{jq}\delta_{kp}.
\end{align}
The covariance matrix of the pure Gaussian state $e^{\theta \hat{c}_p\hat{c}_q}\ket{GS}$ can be written as:
\begin{align} \label{eq:cov_m_rot}
    M_{jk}=\bra{GS}e^{-\theta \hat{c}_p\hat{c}_q}i\hat{c}_j\hat{c}_ke^{\theta\hat{c}_p\hat{c}_q}\ket{GS}-i\delta_{jk}.
\end{align}
Measurement of a Majorana pair can be realized by a projective measurement (we only deal with outcome $+1$ here, because $-1$ outcome can be obtained by measuring $i\hat{c}_q\hat{c}_p$ instead of $i\hat{c}_p\hat{c}_q$). So the covariance matrix of the post-measurement state can be expressed as:
\begin{align} \label{eq:cov_m_measure}
    M_{jk}=\dfrac{\bra{GS}(1+i\hat{c}_p\hat{c}_q)i\hat{c}_j\hat{c}_k(1+i\hat{c}_p\hat{c}_q)\ket{GS}}{2\bra{GS}(1+i\hat{c}_p\hat{c}_q)\ket{GS}},
\end{align}
\roundd{}{if $j\neq k$, and 0 otherwise.}
For further progress we need to express the transformed covariance matrices from the initial ones. As far as we are working with pure Gaussian states we can use Wick's theorem for this purpose,
\begin{align}
    i^p\langle GS|\hat{c}_{j_1}\hat{c}_{j_2}...\hat{c}_{j_{2p}}|GS\rangle=\text{Pf}(M(|GS\rangle)_{j_1,j_2,...,j_{2p}}).
\end{align}
Therefore we are able to express the expectation value of any Majorana chain as the Pfaffian of a sub-matrix of the covariance matrix.

\bigbreak

Through lengthy, but quite straightforward calculations one can derive the efficient algorithms, which realize the transformations of the covariance matrix expressed in Eqs.~\eqref{eq:cov_m_rot}, \eqref{eq:cov_m_measure}.

Here we show the algorithm for the coherent rotation around the Z-axis Alg.~\ref{alg:rot}, and the algorithm for the measurement of Majorana pairs Alg.~\ref{alg:mes}

\begin{algorithm}[!ht] 
\caption{Rotation (M,$\theta$,p,q)}
\label{alg:rot}
$M'[p,:]\gets M[p,:] \cos(2\theta)-M[q, :]\sin(2\theta)$ \\
$M'[q,:]\gets M[q,:]\cos(2\theta)+M[p, :]\sin(2\theta)$ \\
$M'[:,p]\gets M[:,p]\cos(2\theta)-M[:,q]\sin(2\theta)$ \\
$M'[:,q]\gets M[:,q]\cos(2\theta)+M[:,p]\sin(2\theta)$ \\
$M'[p,p] \gets 0$ \\
$M'[q,q] \gets 0$ \\
$M'[p,q]\gets M[p,q]\cos^2(2\theta)-M[q,p]\sin^2(2\theta)$ \\
$M'[q,p]\gets -M[p,q]\cos^2(2\theta)+M[q,p]\sin^2(2\theta)$ \\
\Return $M'$ 
\end{algorithm}

\begin{algorithm}[!ht]
\caption{Measurement (M,p,q)}
\label{alg:mes}
probability $\gets (1/2)(1+M[p,q])$ \\
    \If{$p \neq 0$}{
        $\mathbf{K} \gets M[p,:]$ \\
        $\mathbf{L} \gets M[q,:]$ \\
        $M' \gets M + (1/2p)(\mathbf{L}\mathbf{K}^T-\mathbf{K}\mathbf{L}^T) $ \\
        $M'[p,:] \gets 0$ \\
        $M'[q,:] \gets 0$ \\
        $M'[:,p] \gets 0$ \\
        $M'[:,q] \gets 0$ \\
        $M'[p,q] \gets 1$ \\
        $M'[q,p] \gets -1$}
        \Return ($M'$, probability)
\end{algorithm}

\subsection{Simulating error correction}

Now we will write a more detailed description on how to perform the first two steps of our algorithm described in Sec.~(\ref{sec:Simulation_method}).
In the spirit of Eq.~\eqref{eq:inhom_rounds} it is enough to simulate one error correction round for inhomogeneous coherent errors. However sampling from the probability distribution $P(s)$, and calculation of the assigned logical rotation angle $\theta_L(s)$ involve the previously introduced \roundd{FLO}{fermionic linear optics} operations Algs.~\ref{alg:rot},\ref{alg:mes}

\bigbreak

One of the most painful restriction is that we can not simulate stabilizer measurements efficiently because they involve the joint measurement of 8 Majoranas. Therefore instead of measuring X stabilizers we measure X operators of individual C4 codes, and instead of sampling from probability distribution of syndromes, we are sampling from the distribution of X measurement outcomes. We are allowed to do this, because X stabilizer measurements can be led back to X measurements. For this purpose we introduce the concept of $m$ syndrome, which stores the X measurement outcomes.
\begin{align}
    m \in \{+1,-1\}^{n}
\end{align}
We assign an $s$ syndrome to every $m$ syndrome through the following function:
\begin{align}
    s=\delta(m): \quad s_f=\prod_{j\in\partial f}m_j.
\end{align}
We can sample from probability distribution of $m$ syndromes $P(m)$, because
\begin{align}
    P(s)=\sum_{m:\delta(m)=s}P(m).
\end{align}
It is important to note that this is allowed only, because X measurements are commuting just like stabilizer measurements. If we want to measure both kind of (X and Z) stabilizers this simplification to the measurement of individual Pauli operators is prohibited, because of the non-trivial commutation relation. This is the reason behind our limitation of coherent errors to rotations around Z-axis, because in this case Z stabilizer measurements will be trivial, and we don't have to take them into account.

\bigbreak

In the simulation we are sampling probability distribution $P(m)$ with a Monte Carlo algorithm. We order the measurement outcomes $m_1,m_2,...,m_n$, and draw $m_j$ based on its conditional probability:
\begin{align}
    P(m_j|m_1,m_2,...,m_{j-1})=\dfrac{P(m_1,m_2,...,m_j)}{P(m_1,m_2,...,m_{j-1})}.
\end{align}
We define the probability of a syndrome-part $m_A$:
\begin{align}
    P(m_A)=P(m_1,m_2,...,m_j),
\end{align}
where $A$ is the patch including the first $j$ qubit of the code.

We can express this probability $P(m_A)$ by considering the Majorana representation of the logical $+$ state Eq.~\eqref{eq:Majorana_logical_state} (note $P(s)$ is independent of the initial state, so we can choose $\ket{+_L}$ safely). 
\begin{equation}
\begin{aligned}
    P(m_A)=\nu\bra{\Phi_{link}} \bra{+^L_{C4}}\hat{O}_1^{\dagger},...,\hat{O}_j^{\dagger} \\ 
    \hat{O}_j,...,\hat{O}_1\ket{\Phi_{link}} \ket{+^L_{C4}},
\end{aligned}
\end{equation}
Where $\nu$ is just an $A$ dependent normalization factor, $\ket{+^L_{C4}}$ is the encoded $\ket{+}$ state in the logical C4 code and $\hat{O}_j$ operation contains the projection to the C4 subspace of the $j$-th C4 code, the coherent error on the $j$-th qubit and the measurement of $\hat{X}_j$. This operator has the following form:
\begin{align}
    \hat{O}_k=\dfrac{1}{2}(1+m_k\hat{X}_k)e^{i\theta_k\hat{Z}_k}\dfrac{1}{2}(1+\hat{S}^{(k)}).
\end{align}
$\ket{\Phi_{link}}\ket{+^L_{C4}}$ state is stabilized by all the link operators, $\hat{X}^L_{C4}$ and $\hat{S}^L\hat{X}^L_{C4}$ operators of the logical C4 code, so it is clearly a Gaussian state. Furthermore we are able to write the operator $\hat{O}_k$ as it only contains enabled \roundd{FLO}{fermionic linear optics} operations (measurement of Majorana pairs, and coherent Z rotations),
\begin{align}
    \hat{O}_k=\dfrac{1}{2}(1+m_k\hat{S}^{(k)}\hat{X_k})\dfrac{1}{2}(1+m_k\hat{X}_k)e^{i\theta_k\hat{Z}_k}.
\end{align}

Now it is clear that we can sample from the probability distribution $P(m)$ through \roundd{FLO}{fermionic linear optics} simulations with a Monte Carlo algorithm, but we need to calculate the final state for a given syndrome, or at least the logical rotation angle $\theta_L(s)$.  
For this purpose we write $\tan^2(\theta_L(s))$ in the following form:
\begin{align} \label{eq:tantheta}
    \tan^2(\theta_L(s))=\Bigg|\dfrac{\bra{+_L}\hat{Z}^L\ket{\Phi_s}}{\braket{+_L|\Phi_s}}\Bigg|^2.
\end{align}
Starting from the product state $\ket{+}^{\otimes n}$ and measure all Z stabilizers we can get $\ket{+_L}$ state. However since measuring Z stabilizers commute with the measurement of X stabilizers, correction operator and coherent Z errors, we can write Eq.~\eqref{eq:tantheta} as,
\begin{align}
    \tan^2(\theta_L)=\Bigg|\dfrac{\bra{+^{\otimes n}}\hat{Z}^L\ket{\Phi_s}}{\braket{+^{\otimes n}|\Phi_s}}\Bigg|^2.
\end{align}
Finally one can observe that these expressions are just special cases for $P(m)$, because $\ket{+}^{\otimes n}\bra{+}^{\otimes n}=\hat{\Pi}_{m=0}$ is just the projection when all $m_j=1$. So finally we can write
\begin{align} \label{eq:tanfinal}
    \tan^2(\theta_L)=\dfrac{\bra{+_L}\hat{U}_-^{\dagger}\hat{\Pi}_{m=0}\hat{U}_-\ket{+_L}}{\bra{+_L}\hat{U}_+^{\dagger}\hat{\Pi}_{m=0}\hat{U}_+\ket{+_L}},
\end{align}
where $\hat{U}_-=\hat{Z}^L\hat{C}_s\hat{U}$ and $\hat{U}_+=\hat{C}_s\hat{U}$, so these are just special cases of inhomogeneous coherent errors: $\prod_j\exp(i\theta_j\hat{Z}_j)$.

One can also derive the following relation in a similar way:
\begin{align} \label{eq:tanfinal2}
    \tan^2(\theta_L-\pi/4)=\dfrac{\bra{Y_L}\hat{U}_-^{\dagger}\hat{\Pi}_{m=0}\hat{U}_-\ket{Y_L}}{\bra{Y_L}\hat{U}_+^{\dagger}\hat{\Pi}_{m=0}\hat{U}_+\ket{Y_L}},
\end{align}
where $\ket{Y_L}$ is the logical $+1$ eigenstate of $\hat{Y}^L$ operator. With Eqs.~\eqref{eq:tanfinal},\eqref{eq:tanfinal2} in hand we are able to determine the logical rotation angle modulo $\pi$. 

\bigbreak

An additional thing which is not crucial, but it can help you to reduce the cost of the simulation is the simulation only the active part of the covariance matrix. This means that you don't need to store the whole covariance matrix, only the non-trivial ($M_{jk} \neq 1$, $M_{jk} \neq 0$) sub-matrix. This can be done due to the fact that the measurements of Majorana pairs unbuckle rows and columns from the active part of the covariance matrix. Rotations are bringing in some new rows and columns, but the size of the active sub-matrix will be still notably smaller than the full covariance matrix.

\end{document}